\documentclass[aps, prd, superscriptaddress, eqsecnum, nofootinbib, preprintnumbers, showpacs]{revtex4}
\usepackage{graphicx}

\newcommand{\beq}{\begin{equation}}
\newcommand{\eeq}{\end{equation}}
\newcommand{\beqs}{\begin{eqnarray}}
\newcommand{\eeqs}{\end{eqnarray}}
\newcommand{\zm}{z_{m}}
\newcommand{\ep}{\epsilon}
\newcommand{\tr}{\textrm{Tr}\,}
\newcommand{\condense}{\langle \bar{T}T \rangle}
\newcommand{\Tr}{\hbox{Tr}}

\newcommand{\VEV}[1]{\langle #1 \rangle}

\newcommand{\TeV}{\hbox{TeV}}
\newcommand{\GeV}{\hbox{GeV}}
\newcommand{\MeV}{\hbox{MeV}}

\newcommand{\GB}{ \langle \alpha G_{\mu \nu}^2 \rangle}
\newcommand{\siml}{\hspace{0.3em}\raisebox{0.4ex}{$<$}
\hspace{-0.75em}\raisebox{-.7ex}{$\sim$}\hspace{0.3em}}
\newcommand{\simg}{\hspace{0.3em}\raisebox{0.4ex}{$>$}
\hspace{-0.75em}\raisebox{-.7ex}{$\sim$}\hspace{0.3em}}

\begin{document}

 \title{{\bf  Holographic Techni-dilaton~\footnote{
Short report of this paper was given at 
\cite{Haba:2010iv}. }
 \vspace{5mm}}}

 \author{{Kazumoto Haba}}\thanks{
      {\tt Present address:  
MRI Research Associates Inc., 
                    Tokyo 101-0047, Japan. 
E-mail: kazumohaba@gmail.com}}
       \affiliation{Department of Physics, Nagoya University,
Nagoya, 464-8602, Japan.}
\author{{Shinya Matsuzaki}}\thanks{
      {\tt synya@pusan.ac.kr}}
      \affiliation{ Department of Physics, Pusan National University,
                    Busan 609-735, Korea.}
\author{{Koichi Yamawaki}} \thanks{
      {\tt yamawaki@kmi.nagoya-u.ac.jp}}
      \affiliation{ Kobayashi-Maskawa Institute for the Origin of Particles and 
the Universe (KMI) \\ 
 Nagoya University, Nagoya 464-8602, Japan.}

\date{\today}

\begin{abstract}
Techni-dilaton, a pseudo-Nambu-Goldstone boson of  scale symmetry, was predicted long ago in the 
Scale-invariant/Walking/Conformal Technicolor (SWC-TC) as a remnant of the (approximate) scale symmetry associated with the conformal fixed point, based on the conformal gauge dynamics of ladder Schwinger-Dyson (SD) equation with non-running coupling. We study the techni-dilaton as a flavor-singlet  bound state of techni-fermions by including the techni-gluon condensate (tGC) effect
into the previous (bottom-up) holographic approach to the SWC-TC, a deformation of the holographic QCD with
$\gamma_m \simeq 0$ by large anomalous dimension $\gamma_m \simeq 1$. With including a bulk scalar field corresponding to the gluon condensate, we first improve the Operator Product Expansion of the current correlators so as to reproduce gluonic  $1/Q^4$ term both in QCD and SWC-TC.  We find in QCD about 
$10\%$ (negative) contribution of gluon condensate to the $\rho$ meson mass. We also calculate
the oblique electroweak $S$-parameter in the presence of 
the effect of the tGC and find that for the fixed value of  $S$
the tGC effects dramatically reduce 
the flavor-singlet scalar (techni-dilaton) mass $M_{\rm TD}$ (in the unit of $F_\pi$), while the vector and axial-vector masses
$M_\rho$ and $M_{a_1}$ are rather insensitive to the tGC, where $F_\pi$ is the decay constant of the 
techni-pion. 
If we use the range of values of tGC implied by the 
ladder SD analysis of the non-perturbative scale anomaly in the large $N_f$ QCD near the conformal 
window, the phenomenological constraint 
$S \simeq 0.1$ predicts 
the techni-dilaton mass $M_{\rm TD} \sim 600$ GeV which is within reach of LHC discovery.
\end{abstract}
\maketitle

\section{Introduction}

The origin of mass is the most urgent issue of the particle physics today and is 
to be resolved at the LHC experiments. 
In the standard model,  all masses are attributed to a single parameter of the vacuum expectation value (VEV) of 
the hypothetical elementary particle, the Higgs boson, which is simply transferred  from the input mass parameter $\mu$ 
tuned as tachyonic ($\mu^2<0$) in an ad hoc manner. 
As such the standard model does not explain the origin of mass.

 Technicolor (TC)~\cite{Weinberg:1975gm} is an attractive idea to account for the origin of mass 
 without introducing Higgs boson and tachyonic mass parameter in analogy with QCD: The mass arises dynamically from the condensate of the fermion-antifermion pair
 like Cooper pair in the Bardeen-Cooper-Schrieffer theory of the superconductor, picking up the intrinsic mass scale $\Lambda_{\rm TC}$ generated by the scale anomaly~\cite{Adler:1976zt} through quantum effects in the gauge theory which is scale-invariant at classical level for massless flavors.
 Actually in QCD,  the running of the coupling constant
 $\alpha(\mu)$ implies existence of the intrinsic mass scale $\Lambda_{\rm QCD}$. 
 The original version of TC, 
just a simple scale-up 
of QCD, however,  is plagued by the notorious problems: Excessive flavor-changing neutral currents (FCNCs), and
excessive oblique corrections of 
${\cal O}(1)$ to the Peskin-Takeuchi $S$ parameter~\cite{Peskin:1990zt} compared with the typical
experimental bound about 0.1.

The FCNC problem was resolved long time ago by the scale-invariant/walking/conformal TC 
(SWC-TC)~\cite{Yamawaki:1985zg, Akiba:1985rr, Appelquist:1986an}\cite{Holdom:1984sk}
initially dubbed as ``scale-invariant TC" with the prediction of a ``techni-dilaton'', a pseudo Nambu-Goldstone boson 
of the spontaneous breaking of the (approximate) scale invariance~\cite{Yamawaki:1985zg}. The theory was
 based on the 
strong coupling solution of Maskawa-Nakajima~\cite{Maskawa:1974vs} for the ladder Schwinger-Dyson (SD) equation with {\it non-running} (conformal) gauge coupling $\alpha>\alpha_{\rm cr}={\cal O} (1)$. 
It was found~\cite{Yamawaki:1985zg} (for reviews, see \cite{Miransky:vk}) that the solution implies a large anomalous dimension $\gamma_m \simeq 1$  
of the techni-fermion condensate operator
$\bar T T$ at $\alpha \simeq \alpha_{\rm cr}$,  
namely the enhanced condensate 
$\frac{\langle \bar{T} T \rangle_\Lambda}{\langle \bar{T} T \rangle_m} 
= (\frac{\Lambda}{m})^{\gamma_m}
\simg \frac{\Lambda}{m}$ 
to avoid the FCNC problem, 
where  $m$ is the dynamical mass of techni-fermion on the order of weak scale $\simeq 250$ GeV and the cutoff $\Lambda$ is usually identified with the scale of Extended Technicolor (ETC)~\cite{Dimopoulos:1979es}, 
$\Lambda=\Lambda_{\rm ETC}$ ($>10^3 m$). 
In contrast to a folklore that TC is a ``higgsless model'', a salient feature of SWC-TC  is the conformality which manifests itself  by the appearance of  a composite Higgs boson (``conformal Higgs'') as the techni-dilaton proposed initially in reference~\cite{Yamawaki:1985zg}. 
It is this (approximate) conformal symmetry that is responsible for the naturalness of the SWC-TC to guarantee the large hierarchy $m\ll \Lambda$ in such a way that 
the coupling is almost non-running (conformal) over the wide energy range $m<\mu<\Lambda$.
Moreover, there also exists a possibility~\cite{Appelquist:1991is,Harada:2005ru} that the $S$ parameter can be reduced in the case of SWC-TC.

An explicit gauge dynamics~\cite{Appelquist:1996dq, Miransky:1996pd}  of such an SWC-TC is 
based on the Caswell-Banks-Zaks infrared fixed point~\cite{Caswell:1974gg,Banks:1981nn} (CBZ-IRFP)  $\alpha_* =\alpha_*(N_c,N_f)$  which appears in the two-loop
beta function of  the ``large $N_f$ QCD'', QCD with the number of massless flavors $N_f \, ( < 11N_c/2)$ larger than a certain value  
$N_f^* (\gg N_c)$. 
Note that  $ \alpha_* \searrow 0$ when $N_f  \nearrow 11 N_c/2$, 
and hence there exists a certain region $(N_f^* <) \, N_f^{\rm cr}< N_f < 11 N_c/2 $ (``conformal window'') 
such that   $ \alpha_*  < \alpha_{\rm cr}$,  
where $\alpha_{\rm cr} $ is the critical coupling for the spontaneous chiral symmetry breaking 
and hence the chiral symmetry gets restored in this region.
Here $\alpha_{\rm cr} $ may be evaluated as 
$\alpha_{\rm cr} =\pi/3 C_2(F)$ in the ladder approximation~\cite{Fukuda:1976zb}, 
in which case  $\alpha_*=\alpha_*(N_c,N_f^{\rm cr}) = \alpha_{\rm cr}$ determines  $N_f^{\rm cr} $ as  $N_f^{\rm cr} \simeq 4 N_c$~\cite{Appelquist:1996dq} 
\footnote{
In the case of  $N_c=3$, this value $N_f^{\rm cr} \simeq 4 N_c =12$ is somewhat different 
from the lattice value~\cite{Iwasaki:2003de} \ $6<N_f^{\rm cr}<7$, but is consistent with more recent lattice results~\cite{Appelquist:2007hu}.
} \
\footnote{
There is another possibility for the SWC-TC with much less $N_f$ based on the higher TC representation~\cite{Hong:2004td},
although explicit ETC model building would be somewhat involved.}. 
Related to the conformal symmetry, this phase transition (``conformal phase transition"\cite{Miransky:1996pd}) has 
unusual nature that the order parameter changes continuously but the spectrum does discontinuously 
at the phase transition point $\alpha_*=\alpha_{\rm cr}$  when we change  $\alpha_*$ (or $N_f/N_c$) continuously.

\begin{figure}[h]
\begin{center}
 \includegraphics[width=6cm]{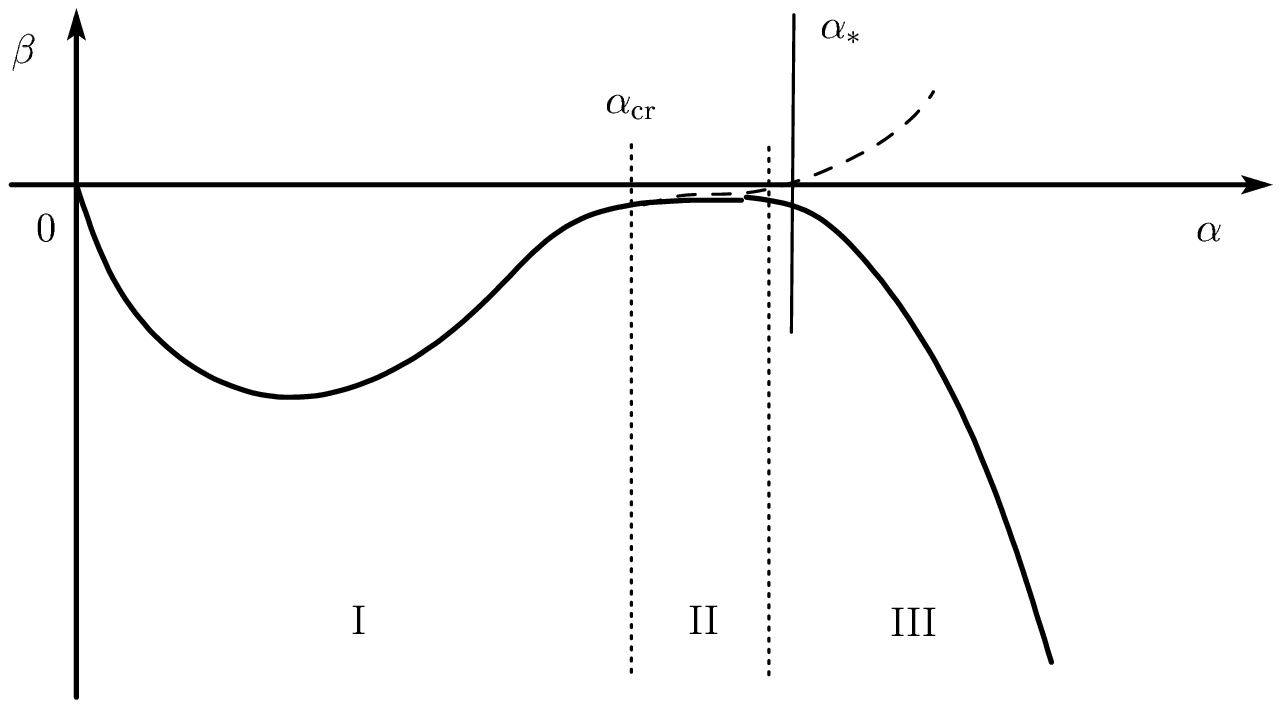}
 \includegraphics[width=7cm]{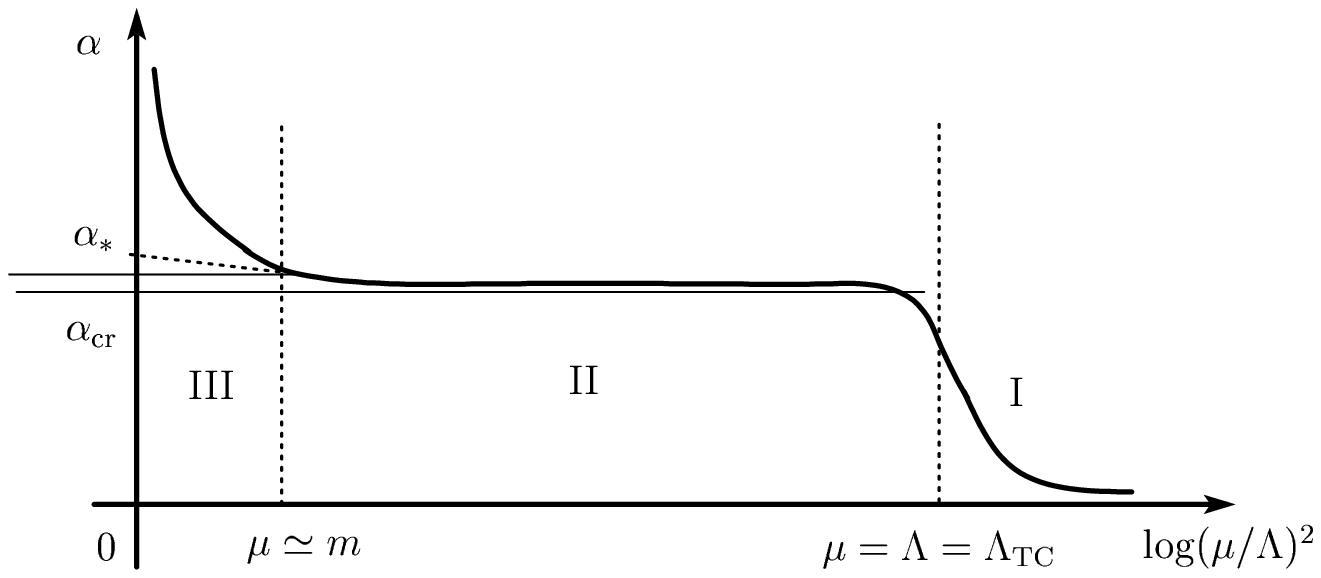} 
\caption{\label{fig:beta}
The beta function and $\alpha(\mu)$ for SWC-TC.}
\end{center} 
 \end{figure}

When we set $\alpha_*$ slightly larger than $\alpha_{\rm cr}$ 
(slightly outside of the conformal window),  the walking 
coupling $\alpha(\mu) (<\alpha_*)$ becomes larger than the critical coupling 
in the wide infrared region,  
we have a condensate or the dynamical mass of the techni-fermion $m$, 
which is much smaller than
the intrinsic scale of the theory $\Lambda_{\rm TC} (\gg m )$. Such an intrinsic scale 
$\Lambda_{\rm TC}$ is quantum mechanically induced by the scale anomaly  as  an analogue of
$\Lambda_{\rm QCD}$  in QCD and the theory behaves as ordinary asymptotically 
free theory as QCD for $\mu > \Lambda_{\rm TC}$ (Region I in Fig.~\ref{fig:beta}). 
Although the  CBZ-IRFP $\alpha_*$ actually disappears (then would-be IRFP) 
at the scale $\mu \sim m$ where the fermions have acquired 
the mass $m$ and get decoupled from the beta function for $\mu<m$ (Region III in Fig.~\ref{fig:beta}),  the coupling is 
still walking due to the remnant of the CBZ-IRFP conformality 
in a wide region $m < \mu < \Lambda_{\rm TC} $ (Region II in Fig.~\ref{fig:beta}). 
Then the theory acts like an SWC-TC: 
$\Lambda_{\rm TC}$ plays a role of 
cutoff $\Lambda$  identified with the ETC scale: $\Lambda_{\rm TC} =\Lambda=\Lambda_{\rm ETC}$. 
It develops a large anomalous dimension $\gamma_m \simeq 1$ for Region II 
to solve the FCNC problem~\cite{Appelquist:1996dq, Miransky:1996pd}.

Existence of these two largely separated scales, 
$m$ and $\Lambda_{\rm TC}$ such that $m\ll \Lambda_{\rm TC}$,  
is the most important feature of SWC-TC,  in sharp contrast to 
the ordinary QCD with small number of flavors (in the chiral limit)  
where all the mass parameters like dynamical mass of quarks are of
order of the single scale parameter of the theory $\Lambda_{\rm QCD}$, $m \sim \Lambda_{\rm QCD}$. 
The intrinsic scale $\Lambda_{\rm TC}$ is related with the scale anomaly corresponding to
the {\it perturbative} running effects
of the coupling, with the ordinary  beta function 
$\beta (\alpha)$ in the Region I, 
in the same sense as in QCD~\cite{Adler:1976zt}.
\beqs
\VEV{\partial^\mu D_\mu} 
= \VEV{\theta^{\mu}_{\mu}} 
=4 \VEV{ \theta_{0}^{0}}
= \frac{\beta (\alpha)}{4 \alpha^2} \GB ={\cal O} (\Lambda_{\rm TC}^4), 
\eeqs
which implies that all the techni-glue balls have mass of ${\cal O} (\Lambda_{\rm TC})$.
On the other hand, the scale $m$ is related with totally different scale anomaly due to the dynamical generation of $m$ which does exist 
even in the idealized case with non-running coupling $\alpha(\mu) \equiv \alpha ( >\alpha_{\rm cr})$ such as the Maskawa-Nakajima solution~\cite{Maskawa:1974vs}, as was discussed some time ago~\cite{Miransky:1989qc}. 
Such an idealized case well simulates the dynamics of Region II~\cite{Appelquist:1996dq, Miransky:1996pd}, with anomalous dimension $\gamma_m \simeq 1$ and $m \ll \Lambda_{\rm TC}$ 
in the numerical calculations~\cite{Harada:2003dc} with the 
walking coupling constant  in the most of the Region II being slightly larger 
than $\alpha_{\rm cr}$, $ \alpha_* >\alpha(\mu) > \alpha_{\rm cr}$.  
The coupling $\alpha \equiv \alpha_*$ in the ``idealized Region II" 
actually runs non-perturbatively according to the essential-singularity scaling 
(Miransky scaling~\cite{Miransky:1984ef}) of mass generation, 
 with the {\it non-perturbative} beta function $\beta_{\rm NP}(\alpha)$.  
Then the {\it non-perturbative} scale anomaly reads~\cite{Miransky:1996pd} 
\beqs
\VEV{\partial^\mu D_\mu} 
= \VEV{\theta^{\mu}_{\mu}} 
=4 \VEV{ \theta_{0}^{0}}
= \frac{\beta_{\rm NP}(\alpha)}{4 \alpha^2} \GB ={\cal O}(m^4), 
\eeqs
which vanishes when we approaches the conformal window from the broken phase $\alpha_* \searrow \alpha_{\rm cr}$ ($m \to 0$) where $\alpha_*$ is the would-be CBZ-IRFP (See 
Eqs.(\ref{pD}) and (\ref{beta})). 
All the techni-fermion bound states have mass of order of $m$, while there are no light bound states
in the symmetric phase (conformal window) $\alpha_*<\alpha_{\rm cr}$, a characteristic feature of the conformal phase transition~\cite{Miransky:1996pd}.
The techni-dilaton is associated with the latter scale anomaly and should have mass on order of $m (\ll \Lambda_{\rm TC})$.

More concretely, the mass of techni-dilaton or scalar bound state in the SWC-TC was estimated in various methods:   
The first method was based on the assumption of partially conserved dilatation current (PCDC)~\cite{Shuto:1989te}  combined with the  ladder SD equation for the gauged Nambu-Jona-Lasinio model which well simulates~\cite{Appelquist:1996dq,Miransky:1996pd} the 
conformal phase transition in the large $N_f$ QCD. The result indicates
\beqs
M_{\rm TD} \simeq 
\sqrt{2} m \, ,
\label{PCDCmass}
\eeqs
which  coincides with other methods in the ladder SD equation without use of the PCDC~\cite{Carena:1992cg}.
Also a straightforward calculation~\cite{Harada:2003dc,Kurachi:2006ej} of scalar bound state mass as well as the $S$ parameter~\cite{Harada:2005ru} was made  in the
vicinity of the CBZ-IRFP in the large $N_f$ QCD, based on the coupled use of the SD equation and the Bethe-Salpeter (BS) equation in the ladder approximation: 
\beqs
M_{\rm TD} \sim 1.5\, m \,  (< M_\rho, M_{a_1})\, ,
\eeqs
where  $M_\rho, M_{a_1}$ are masses of (techni-)$\rho$ and (techni-)$a_1$ mesons, respectively, which is consistent with Eq.(\ref{PCDCmass}) in contrast to the ordinary QCD where the scalar mass is larger than those of the
 vector mesons (``higgsless'') within the same framework of ladder
 SD/BS equation approach.

The SWC-TC,  
however, has a calculability problem, since its 
non-perturbative dynamics is not QCD-like at all, 
and hence no simple scaling of QCD results would be available. 
The best thing we could do so far has been a straightforward calculation 
based on the SD equation and (inhomogeneous) BS equation 
in the ladder approximation~\cite{Harada:2005ru}, 
which is however not a systematic approximation and is not very reliable in the quantitative sense.

   Of a late fashion, based on the so-called AdS/CFT (
anti-de-Sitter space/conformal field theory) correspondence,  
a duality of the string in the anti-de Sitter space background-conformal 
field theory~\cite{Maldacena:1997re}, 
holography gives us a new method which may resolve 
the calculability problem of strongly coupled gauge theories~\cite{Arkani-Hamed:2000ds}: 
Use of the holographic correspondence enables us to 
calculate Green functions in a four-dimensional strongly coupled theory 
from a five-dimensional weakly coupled theory. 
For instance, QCD can be reformulated based on the holographic correspondence 
either in the bottom-up approach~\cite{DaRold:2005zs,Erlich:2005qh}  
or in the top-down approach~\cite{Sakai:2004cn}. 
In both approaches we end up with the five-dimensional gauge theory for the flavor symmetry, 
whose infinite tower of Kaluza-Klein (KK) modes describe nicely a set of the massive vector/axial-vector mesons 
as the gauge bosons of Hidden Local Symmetries (HLSs)~\cite{Bando:1984ej,Harada:2003jx}, 
or equivalently as the Moose~\cite{Georgi:1985hf}. 
Although a holographic description is valid only for large $N_c$ limit, 
several observables of QCD have been reproduced 
within 30~\% errors in both approaches. Moreover,  
through the high-energy behavior of current correlators  in 
operator product expansion (OPE), some consistency with the QCD
has been confirmed in the bottom-up approach.

Recently the $S$ parameter in the SWC-TC was calculated~~\cite{Hong:2006si,Piai:2006hy,Agashe:2007mc,Haba:2008nz}  
as an application of the above technique of 
bottom-up holographic QCD (hard-wall model)~\cite{DaRold:2005zs,Erlich:2005qh} 
to the holographic SWC-TC. 
   In the previous work~\cite{Haba:2008nz}, 
based on the holographic correspondence in the bottom-up approach, 
we calculated the $S$ parameter in the SWC-TC, 
treating the anomalous dimension $\gamma_m$ 
as a free parameter as $0 \siml \gamma_m  \siml 1$, varying continuously 
from the QCD monitor value  $\gamma_m \simeq 0$ 
through the one of the SWC-TC  $\gamma_m \simeq 1$. 
We obtained $S$ as an explicit function of $F_\pi/M_\rho$ in 
entire region, which turns out to be a positive and monotonically increasing 
function such that $S$ continuously goes to zero when $F_\pi/M_\rho \to 0$, 
where $F_\pi$ and $M_\rho$ are the (techni-)pion decay constant and 
the (techni-)$\rho$ meson mass, respectively.

In this paper, 
we extend the previous paper~\cite{Haba:2008nz} on the hard-wall-type 
bottom-up holographic SWC-TC by including effects of (techni-) gluon 
condensation, $\Gamma$, through the bulk flavor/chiral-singlet scalar field $\Phi_X$, 
in addition to the conventional bulk scalar field $\Phi$ dual to the chiral condensate. 
 For definition of $\Gamma$, see text. 
The techni-dilaton, a flavor-singlet scalar bound state of
techni-fermion and anti-techni-fermion, will be identified with the lowest KK mode 
coming from the bulk scalar field $\Phi$, not $\Phi_X$.   
Thanks to the additional explicit bulk scalar field $\Phi_X$, 
we naturally improve the matching with the OPE 
of the underlying theory (QCD and SWC-TC) for current correlators 
so as to reproduce gluonic $1/Q^4$ term, 
which is clearly distinguished from the same $1/Q^4$ terms from 
chiral condensate in the case of SWC-TC with $\gamma_m \simeq 1$. 
Our model with $\gamma_m =0$ and $N_f=3$ 
well reproduces the real-life QCD (See Table~\ref{table:QCD}). 
We find that the QCD $\rho$ meson mass $M_\rho$ includes 
a (negative) contribution about 10\% from the gluon condensate.

We analyze a generic case with $0 \lesssim \gamma_m \lesssim 1$ 
and calculate masses of the techni-$\rho$ meson ($M_\rho$), 
the techni-$a_1$ meson ($M_{a_1}$), 
and the flavor-singlet scalar meson, techni-dilaton ($M_{\rm TD}$), 
as well as the $S$ parameter. 
We discuss the general tendency of the dependence of 
the meson masses relative to $F_\pi$, 
($M_\rho/F_\pi,\,M_{a_1}/F_\pi,\,M_{\rm TD}/F_\pi$) on $\gamma_m$, $S$, 
and $\Gamma$. 
  We find a characteristic feature of the techni-dilaton mass related to 
the conformality of SWC-TC:   
For fixed $S$ and $\gamma_m$, 
$(M_\rho/F_\pi)$ and $(M_{a_1}/F_\pi)$ are not sensitive to $\Gamma$, 
while $(M_{\rm TD}/F_\pi)$ substantially decreases as $\Gamma$ increases. 
Actually, in the formal  limit $\Gamma \to \infty$, we would have $M_{\rm TD}/F_\pi  \to 0$ 
(This does not imply the existence of the isolated true massless NG boson of the scale symmetry, 
since in our case the decay constant $F_{\rm TD}$ diverges and the techni-dilaton gets decoupled in that limit, see text.) 
  For fixed $S$ and $\Gamma$, 
again $(M_\rho/F_\pi)$ and $(M_{a_1}/F_\pi)$ are not sensitive to $\gamma_m$,
while $(M_{\rm TD}/F_\pi)$ substantially decreases as $\gamma_m$ increases. 

 Particularly for the case of $\gamma_m = 1$, 
we study the dependence of the $S$ parameter on $(M_\rho/F_\pi)$ 
for typical values of $\Gamma$. 
It is shown that the techni-gluon contribution reduces the value of $S$ 
maximally about 10\% in the region of $\hat S \lesssim 0.1$, 
although the general tendency is similar to 
the previous paper~\cite{Haba:2008nz} without techni-gluon condensation:  
$\hat{S}$ decreases monotonically with respect to $(F_\pi/M_\rho)$ to continuously approach zero. 
This implies $(M_\rho/F_\pi)$ necessarily increases when $\hat S$ is required
to be smaller.

To be more concrete, we consider a couple of typical models of SWC-TC 
with $\gamma_m \simeq1$ and $N_{\rm TC} = 2,3,4$ based on the CBZ-IRFP  
in the large $N_f$ QCD. 
Using some specific dynamical features of the conformal anomaly 
indicated by the analysis based on the ladder SD equation (Eqs.(\ref{pD}) and (\ref{beta})),
we find the relation of $\Gamma$ to $(\Lambda_{\rm ETC}/F_\pi)$: 
In the case of $N_{\rm TC}=3$ ($N_f = 4 N_{\rm TC}$) and $S \simeq 0.1$,
we have $\Gamma \simeq 7$ for $(\Lambda_{\rm ETC}/F_\pi) = 10^4$--$10^5$
(required by the FCNC constraint).
Thanks to the large anomalous dimension $\gamma_m$ and 
large techni-gluon condensation $\Gamma$, we obtain 
a relatively light techni-dilaton with mass $M_{\rm TD} \simeq 600 \, \GeV$
compared with $M_\rho \simeq M_{a_1} \simeq 3.8 \, \TeV$,  
consistently with the perturbative unitarity of $W_LW_L$ scattering. 
Note that $M_\rho$ and $M_{a_1}$ are essentially determined 
by the requirement of $S = 0.1$ fairly independently of techni-gluon 
condensation. 
The essential reason for the large $\Gamma$ 
is due to the existence of 
the wide conformal region $F_\pi < \mu < \Lambda_{\rm ETC}$
with $(\Lambda_{\rm ETC}/F_\pi) = 10^4$--$10^5$,
which yields the smallness of the beta function 
through the factor $(\ln{4\Lambda_{\rm ETC}/m})^{-3}$
(see Eq.({\ref{beta})) and hence amplifies 
the techni-gluon condensation 
compared with the ordinary QCD with $\Gamma=1$. 
In the idealized (phenomenologically non-interesting) limit $\Lambda_{\rm ETC}/F_\pi \to \infty$ 
we would have $\Gamma \to \infty$ and hence $M_{\rm TD}/F_\pi  \to 0$ in conformity with the general tendency mentioned above.  
(However, the would-be  ``massless'' techni-dilaton is actually decoupled, see text. 
Indeed,  spontaneous breaking of the scale symmetry is always accompanied with its explicit breaking.)
The predicted mass $\simeq 600$ GeV of the holographic techni-dilaton 
(``conformal Higgs") is within reach of LHC discovery.

  This paper is organized as follows: 
In Sec.~\ref{review} we present our model which is an extension 
of our previous holographic SWC-TC model based on 
hard-wall-type bottom-up approach~\cite{Haba:2008nz} 
by including effects of (techni-) gluon 
condensation through the bulk flavor/chiral-singlet scalar field.  
 Formulas for masses of mesons (techni-$\rho$, -$a_1$, and -dilaton) and current correlators including $S$ parameter are given.  
We show that our model reproduces gluonic $1/Q^4$ terms in the OPE of 
vector/axial-vector current correlators.  
In Sec.~\ref{sec3} 
we estimate effects on meson masses and $S$ parameter coming from 
the (techni-)gluon condensation $\Gamma$ as a free parameter 
in a generic TC with $0 \lesssim \gamma_m \lesssim 1$ involving the case of QCD with $\gamma_m\simeq 0$. 
In Sec.~IV, 
to specify the value of $\Gamma$ relevant to the 
actual model-building of SWC-TC, 
we consider a matching of our holographic model 
with typical models of SWC-TC based on the CBZ-IRFP 
in the large $N_f$ QCD.

\section{A holographic technicolor model with techni-gluon condensation} 
\label{review}

In this section, we propose a holographic model dual to a generic class of technicolor (TC) 
with $0 \lesssim \gamma_m \lesssim 1$ including the degree of freedom of (techni-)gluon condensation, 
where $\gamma_m$ denotes the anomalous 
dimension of (techni-)fermion chiral condensate $\condense$.

 Following a bottom-up approach of holographic-dual of QCD~\cite{DaRold:2005zs,Erlich:2005qh} 
 with $\gamma_m\simeq 0$ and that of SWC-TC~\cite{Hong:2006si,Piai:2006hy,Haba:2008nz} with $\gamma_m \simeq 1$,  
we consider a five-dimensional gauge theory having 
$SU(N_f)_L \times SU(N_f)_R$ gauge symmetry.
We will not consider 
the extra $U(1)_{A}$ that involves the anomaly.
The theory is defined on the five-dimensional 
anti-de-Sitter space (AdS$_5$) with $L$, the curvature radius of AdS$_5$, 
described by the metric $ds^2= g_{MN} dx^M dx^N 
= 
\left(L/z \right)^2\big(\eta_{\mu\nu}dx^\mu dx^\nu-dz^2\big)$ 
with $\eta_{\mu\nu}={\rm diag}[1, -1, -1,-1]$. 
The fifth direction $z$ 
is compactified on an interval extended from the ultraviolet (UV) brane located at $z=\epsilon$ 
to the infrared (IR) brane at $z=z_m$, i.e., $ \epsilon \leq z \leq z_m  $.  
  In addition to the bulk left- ($L_M$) and right- ($R_M$) gauge fields,  
we introduce a bulk scalar field $\Phi$ which transforms as bifundamental representation under 
the $SU(N_f)_L \times SU(N_f)_R$ gauge symmetry 
so as to deduce the information concerning 
the chiral condensation-operator $\bar{T} T$. 
The mass-parameter $m_\Phi$ is then related to $\gamma_m$ 
as $m_\Phi^2=- (3-\gamma_m)(1+ \gamma_m)/L^2$, 
where $\gamma_m = 0$ corresponds to QCD and QCD-like TC and $\gamma_m \simeq 1$ is the case of SWC-TC. 
This is the same setup as in Refs.~\cite{Hong:2006si,Piai:2006hy,Haba:2008nz}.

  In order to incorporate effects from techni-gluon condensation,  
here we introduce an additional bulk scalar field $\Phi_X$ 
dual to techni-gluon condensate $\langle \alpha G_{\mu\nu}^2
\rangle$, where $\alpha$ is related to the TC gauge couping $g_{\rm TC}$ by $\alpha = g_{\rm TC}^2/(4\pi)$. 
Since $\langle \alpha G_{\mu\nu}^2 \rangle$ is singlet under 
the chiral $SU(N_f)_L \times SU(N_f)_R$ symmetry and $U(1)_V$ symmetry, 
the dual-bulk scalar field $\Phi_X$ has to be a real field. 
We take ${\rm dim}(\alpha G_{\mu\nu}^2)=4$ and 
the corresponding bulk-mass parameter $m_{\Phi_X}^2=0$.

The form of interaction terms involving the new bulk field $\Phi_X$
 still remains undetermined. 
In the present work, we shall adopt a ``dilaton-like'' coupling, 
such that all the fields couple to $\Phi_X$ 
in the exponential form like $e^{\Phi_X(z)}$
\footnote{This should not be confused with 
so-called soft-wall model~\cite{Karch:2006pv}
where $1/z_m=0$ in contrast to our case with $1/z_m \neq 0$.}
. 
($\Phi_X$ is {\it not} identified with techni-dilaton in this paper.)

Thus the five-dimensional action employed in the present paper 
takes the form: 
\beqs
S_5 &=&\,\int\,d^4 x\,\int_{\ep}^{\zm}\,d\,z~\sqrt{-g}\,
\frac{1}{ g_5^2}e^{c g_5^2 \Phi_X(z)} \Big(
-\frac{1}{4}\tr\left[{L_{MN}L^{MN}}
+{R_{MN}R^{MN}}\right]
\nonumber \\ 
&& 
\hspace{150pt} 
+\tr\left[{D_M\Phi^\dagger D^M\Phi}
-m^2_\Phi \Phi^\dagger \Phi \right]
+ \frac{1}{2} \partial_M 
\Phi_X \partial^M \Phi_X \Big)
\,, 
\label{S5}
\eeqs
where $L_M(R_M)=L_M^a(R_M^a) T^a$ with the generators of $SU(N_f)$ normalized by 
${\rm Tr}[T^a T^b]=\delta^{ab}$; 
$L(R)_{MN} = \partial_M L(R)_N - \partial_N L(R)_M 
 - i [ L(R)_M, L(R)_N ]$; 
$g={\rm det}[g_{MN}]=-(L/z)^{10}$; 
$g_5$ denotes the gauge coupling in five-dimension 
and $c$ is the dimensionless coupling constant. 
The covariant derivative acting on $\Phi$ is defined as 
$D_M\Phi=\partial_M \Phi+iL_M\Phi-i\Phi R_M$.

We parametrize the bulk scalar fields $\Phi$ and $\Phi_X$ as follows: 
\beqs
\Phi(x,z) &=& \frac{1}{\sqrt{2}}(v(z)+ \sigma(x,z)) \exp[{i \pi(x,z)/v(z)}]
\,, \label{Phi:decompose} \\
\Phi_X(z) &=& 
v_X(z) 
\,,
\label{PhiX:decompose}
\eeqs
with the vacuum expectation values (VEVs), $v(z)=\sqrt{2} \langle \Phi \rangle$ and 
 $v_X(z)= \langle \Phi_X \rangle$, respectively.  
In Eq.(\ref{PhiX:decompose}) we ignored Kaluza-Klein (KK) modes 
of $\Phi_X$ (including the lowest mode) which are identified with massive glueballs 
with mass of order  
${\cal{O}}(\Lambda_{\rm TC})$ which is much larger than the electroweak scale, $\Lambda_{\rm TC} \gg F_\pi$, 
in the case of SWC-TC with $\gamma_m \simeq 1$. 
The techni-dilaton, a flavor-singlet scalar bound state of
techni-fermion and anti-techni-fermion, will be identified with the lowest KK mode 
of $\sigma(x,z)$, but not of $\Phi_X$.

We choose the boundary condition for $v(z)$ as 
\begin{eqnarray}
\alpha M &=& 
\lim_{\epsilon \to 0}
Z_m 
\left(\frac{L}{z} v(z)\right)
\,\Big|_{z=\epsilon},\qquad 
Z_m = Z_m \left(L/z\right) = \left(\frac{L}{z}\right)^{\gamma_m}, 
\label{M:def}\\
\xi &=& L v(z) \,\Big|_{z=z_m} \, , 
\label{xi:def}
\end{eqnarray}
where $M$ stands for the current mass of techni-fermions and 
$\xi$ is related to the techni-fermion condensate $\langle \bar{T}T \rangle$ 
as will be clarified later (See Eq.(\ref{condense:sol})). 
The parameter $\alpha$ has been introduced which can arise 
from ambiguity of definition of the current mass $M$. 
Here we take $\alpha = \sqrt{3}$, which turns out to be consistent with 
the operator product expansion (OPE) for the scalar current correlator in QCD~\cite{DaRold:2005vr}.

For $v_X$, we impose the following boundary condition:  
\begin{eqnarray}
M' = 
\lim_{\epsilon \to 0}L v_X(z)
\,\Big|_{z=\epsilon}, \qquad
{\cal{G}} = L v_X(z) \,\Big|_{z=z_m} \, , 
\label{vXBC}
\end{eqnarray} 
where $M'$ becomes the external source for the techni-gluon 
condensation-operator $\alpha G^2_{\mu\nu}$ 
and ${\cal G}$ is associated to the techni-gluon condensate 
$ \langle \alpha  G_{\mu\nu}^2 \rangle$ 
as we will see later (See Eq.(\ref{GB:sol})). 
We define the techni-gluon condensate in such a way 
that it does not include a trivial perturbative contribution, namely, 
\begin{equation} 
\langle \alpha G_{\mu\nu}^2 \rangle 
\equiv \langle \alpha G_{\mu\nu}^2 \rangle_{\rm full} 
- \langle \alpha G_{\mu\nu}^2 \rangle_{\rm perturbation}
\,. \label{G2:def} 
\end{equation} 
We then see that ${\cal G}$ in Eq.(\ref{vXBC}) is 
related only to the non-perturbative breaking of the conformal/scale invariance, 
while $M'$ in Eq.(\ref{vXBC}) serves as its source and itself plays a role of the 
explicit breaking of the conformal symmetry just like the current mass $M$ in the case of the chiral symmetry.

We next introduce 
the five-dimensional vector and axial-vector gauge fields 
$V_M$ and $A_M$ defined by 
$V_M=(L_M+R_M)/\sqrt{2} $ and  $A_M= (L_M-R_M)/\sqrt{2}  $. 
The UV boundary values of $V_\mu$ and $A_\mu$ then 
play the role of the sources ($v_\mu$, $a_\mu$) 
for the vector and the axial-vector currents 
externally coupled to TC sector. 
Working in $V_z = A_z \equiv 0$ gauge, 
we choose their boundary conditions as  
\begin{eqnarray}\label{VAmuBC}
&& \partial_z V_{\mu}(x,z)\big|_{z=z_m} =
\partial_z A_{\mu}(x,z)\big|_{z=z_m}=0 
\,, \nonumber \\ 
&&
V_{\mu}(x,z)\big|_{z=\epsilon}=v_{\mu}(x) 
\,, \qquad 
A_{\mu}(x,z)\big|_{z=\epsilon}=a_{\mu}(x) 
\, .
\end{eqnarray}

Once the boundary conditions (\ref{M:def}), (\ref{xi:def}), (\ref{vXBC}), and (\ref{VAmuBC}) are given, 
the equations of motion for the bulk fields can be solved at the classical level. 
By substituting those solutions into the action (\ref{S5}), 
the effective action $S^{\rm eff}_5$ is expressed as a certain functional of the 
UV boundary values, $M$, $M'$, $v_\mu$, and $a_\mu$, i.e., 
$ S_5^{\rm eff} = S_5^{\rm eff}[M, M', v_\mu, a_\mu]$. 
  From the familiar AdS/CFT dictionary, this $S_5^{\rm eff}$ corresponds 
to the generating functional $W$ in TC 
written in terms of  the external sources $M$, $M'$, $v_\mu$, and $a_\mu$. 
One can then readily calculate the two-point Green functions in the usual way: 
\begin{eqnarray} 
\frac{ \delta^2 W[v_\mu]}{\delta \tilde v^a_\mu(q) \delta  \tilde v^b_\nu(-q)} \Bigg|_{v_\mu=0} 
&=& 
i\int d^4 x \,e^{iq\cdot x} \langle 
J^{a\mu}_V (x) J^{b\nu}_V(0) \rangle = 
-\delta^{ab} \left(\eta^{\mu\nu}-\frac{q^{\mu}q^{\nu}}{q^2}\right)\,  \Pi_{V}(-q^2) 
\, , 
\label{Pi:V} \\ 
\frac{ \delta^2 W[a_\mu]}{\delta \tilde a^a_\mu(q)  \delta \tilde a^b_\nu(-q)} \Bigg|_{a_\mu=0} 
&=& 
i\int d^4 x \,e^{iq\cdot x} \langle 
J^{a\mu}_A (x) J^{b\nu}_A(0) \rangle = 
-\delta^{ab} \left(\eta^{\mu\nu}-\frac{q^{\mu}q^{\nu}}{q^2}\right)\, \Pi_{A}(-q^2) 
\, ,
\label{Pi:A} \\ 
\lim_{\ep \to 0}  i \frac{ i \delta W[M]}{\delta M } \Bigg|_{M=0} 
&=&
\langle \bar{T} T \rangle 
\,, \label{TbarT}\\
\lim_{\ep \to 0}  i \frac{ i \delta W[M']}{\delta M' } \Bigg|_{M'=0} 
&=&
\langle \alpha  G_{\mu\nu}^2 \rangle
\,, \label{aGB}
\end{eqnarray}
where $\tilde{v}^\mu(q)$ and $\tilde{a}^\mu(q)$ respectively 
denote  the Fourier component of $v^\mu(x)$ and $a^\mu(x)$.  
To facilitate the later discussions,  as done in Ref.~\cite{Haba:2008nz},  
we introduce $\hat{S}$ as the $S$ parameter per each techni-fermion doublet,  
$\hat{S} = S/(N_f/2) $, 
which is expressed by the vector and axial-vector current correlators $\Pi_V$ and $\Pi_A$ 
as 
\begin{equation}
\hat{S}=-4\pi\,\frac{d}{ d
Q^2}\left[\Pi_V(Q^2)-\Pi_A(Q^2)\right]_{Q^2=0}
\,, 
\label{Sdif}
\end{equation}
where $Q \equiv \sqrt {- q^2}$. 
We also introduce the (techni-)pion decay constant defined as 
\begin{equation} 
  F_\pi^2 = \Pi_V(0) - \Pi_A(0)  
  \,. \label{fpi:def}
\end{equation}

\subsection{Condensates: $\langle \bar{T}T \rangle$ and 
$ \langle \alpha  G_{\mu\nu}^2 \rangle$} 
\label{condensations}

In the following, evaluating the equations of motion for the bulk fields explicitly, 
we shall present the formulas
for the condensates, $\langle \bar{T} T \rangle$ and $\langle \alpha G_{\mu\nu}^2 \rangle$, 
the $S$ parameter and the decay constant $F_\pi$, and masses of 
vector mesons, axial-vector mesons, and scalar mesons.

The action (\ref{S5}) leads to the following equation of motion for $v_X(z)$:
\beqs
\partial_{z}\left(\frac{1}{z^3}\partial_{z}\chi(z) \right) + 
\frac{(c g_5^2)^2}{4L^2}\chi(z) L^2 \tr \left[-\frac{1}{z^3} (\partial_{z}v(z))^2 
+ \frac{(3-\gamma_m)(1+\gamma_m)}{z^5} v^2(z) \right]=0
\,, 
\label{chiEOM0}
\eeqs
where we have defined 
\beq
\chi(z) = 
\exp{\left[\frac{c}{2} g_5^2 v_X(z) \right]} 
\,. 
\eeq 
The boundary condition for $\Phi_X$ given in Eq.(\ref{vXBC}) is now rewritten 
in terms of $\chi$ as  
\begin{eqnarray}
 \chi(z)\Big|_{z=\epsilon}
 &=& 
\exp{\left[\frac{c}{2} g_5^2 v_X(\ep) \right]}
= 
\exp{\left[\frac{c}{2} \frac{g_5^2}{L} M' \right]}
\,,
\label{chiUVBC}\\
\chi(z)\Big|_{z=z_m} &=& 
\exp{\left[\frac{c}{2} g_5^2v_X(z_m)\right]}
= 
\exp{\left[\frac{c}{2} \frac{g_5^2}{L} {\cal{G}} \right]}
\equiv G+1 
\,. 
\label{chiIRBC}
\end{eqnarray} 
We solve the equation of motion (\ref{chiEOM0}) keeping only the first term of 
the left hand side in Eq.(\ref{chiEOM0}). 
This assumption will be justified later in determining the size of 
$|(c/2) (g_5^2/L)|^2$, which turns out to be $\simeq 10^{-4}$ (See Eq.(\ref{g5-c})). 
Equation (\ref{chiEOM0}) is now easily solved to give the solution 
\beq
\label{chi:sol}
\chi(z) = \exp{\left[\frac{c}{2} g_5^2 v_X(z) \right]}  
= c_1^\chi + c_2^\chi \left(\frac{z}{L}\right)^4,
\eeq
where $c_1^\chi$ and $c_2^\chi$ are determined by 
Eqs.(\ref{chiUVBC}) and (\ref{chiIRBC}) in the limit $\ep \to 0$ as 
\beqs
c_1^\chi = e^{\frac{c}{2} \frac{g_5^2}{L} M' },
\qquad
c_2^\chi = \left(\frac{L}{z_m}\right)^4 \left(G+1 - c_1^\chi \right) 
\,. \label{c1:c2:X}
\eeqs
Note that the solution in Eq.(\ref{chi:sol}) gives rise to the induced metric
for the vector and axial-vector gauge fields (See Eq.(\ref{w})).

  We next turn to the equation of motion for $v(z)$ which is read off from  
the action (\ref{S5}) as follows:  
\beqs
\label{vEOM}
\partial_{z}\left(\frac{1}{z^3}\chi^2(z) 
\partial_{z}v(z) \right) 
+\chi^2(z) 
\frac{(3-\gamma_m)(1+\gamma_m)}{z^5}v(z)=0 \,. 
\eeqs   
Substituting Eq.(\ref{chi:sol}) into Eq.(\ref{vEOM}) and 
taking $M'=0$, 
we find the solution in the limit $\ep \to 0$
\beq
\label{v:sol}
v(z) = 
\frac{1}{1+G\left( \frac{z}{z_m} \right)^4} 
\left(c_1\left(\frac{z}{L}\right)^{\gamma_m+1} 
+ c_2 \left(\frac{z}{L}\right)^{3-\gamma_m} \right) 
\,, 
\eeq
where $c_1$ and $c_2$ are determined by the boundary condition in Eqs.(\ref{M:def}) and (\ref{xi:def}) as 
\beqs 
c_1 &=& \sqrt{3} M 
\,, \label{c1:v} \\ 
c_2 &=& \frac{\xi(1+G)}{L} \left( \frac{L}{z_m} \right)^{3-\gamma_m} - \left(\frac{L}{z_m}\right)^{2(1-\gamma_m)} c_1 
\,. 
\label{c2:v}  
\eeqs 
Note that in Eqs.(\ref{v:sol}) and (\ref{c2:v}) (techni-)gluon
condensation effects are included: 
When $G= 0$ in Eqs.(\ref{v:sol}) and (\ref{c2:v}), 
we get back to the previous results without 
(techni-)gluon-condensation effects~\cite{DaRold:2005zs,Erlich:2005qh,Hong:2006si,Piai:2006hy,Haba:2008nz}.

Putting the classical solutions, Eqs.(\ref{chi:sol}) and (\ref{v:sol}), 
into the action (\ref{S5}), 
we are left with the four-dimensional boundary term which is holographically dual to the 
generating functional $W[M,M']$ in TC,  
\beqs
\label{W:MM'}
W[M, M']
&=&\int d^4x \frac{L^3}{2g_5^2} 
\left[ -\frac{1}{z^3}\frac{4}{(c g_5^2)^2}
\partial_z \chi (z) \cdot  \chi(z) 
-\frac{1}{z^3}\chi^2(z)
\tr\left[\partial_z v(z) \cdot  v(z) 
\right]\right]^{\zm}_{\ep}~.
\eeqs 
Using Eqs.(\ref{TbarT}) and (\ref{aGB}) and  performing the functional derivative with respect to the sources $M$ and $M'$, 
the techni-gluon condensate $\GB$ 
and the techni-fermion condensate 
$\condense$ are respectively 
expressed in terms of the five-dimensional gauge theory as 
\begin{eqnarray}
\GB
&=&-8\frac{L^2}{c g_5^4}\frac{1}{z_m^4}G,
\label{GB:sol}\\
\label{condense:sol}
\condense_{1/L}
&=&-\sqrt{3} \frac{L}{g_5^2}\frac{(3-\gamma_m)}{z_m^3}
\left(1 + G \right)\xi \cdot Z_m^{-1},
\eeqs
where $Z_m= (L/z_m)^{\gamma_m}$.

\subsection{Vector, axial-vector current correlators, decay constant $F_\pi$, and $S$ parameter}  
\label{2B}

Let us next focus on the vector and axial-vector sectors.  
The relevant action under the gauge-fixing $V_z=A_z\equiv 0$ reads  
\beqs
S_5 &\ni&\,-\frac{1}{2 g_5^2}\int\,d^4 x\,\int_{\ep}^{\zm}\,d\,z~
w(z)\Big(
\tr\left[{\frac{1}{2}V_{\mu \nu}V^{\mu \nu}
- \partial_z V_\mu \partial_z V^\mu}
+{\frac{1}{2} A_{\mu \nu}A^{\mu \nu} -\partial_z A_\mu \partial_z A^\mu}\right]
\nonumber \\ 
&& 
\hspace{150pt} 
- 2\left(\frac{L}{z}\right)^2
\tr\left[v^2(z) A_\mu A^\mu\right]
\Big)
\,, 
\label{S5:VA}
\eeqs
where $V (A)_{\mu\nu} = \partial_\mu V(A)_\nu - \partial_\nu V (A)_\mu$
and  
the induced metric $w(z)$ is given by the solution in Eq.(\ref{chi:sol}) as
\beq
w(z) = \frac{L}{z} \chi^2(z) 
= \frac{L}{z} \left(1+G \left(\frac{z}{z_m}\right)^4\right)^2
\,. \label{w}
\eeq
In arriving at the last equality we have used Eqs.(\ref{chi:sol}), (\ref{c1:c2:X}),   
and set $M'=0$. (When $M'=0$ and $\ep = 0$ the explicit breaking of conformal/scale invariance
only comes from $1/z_m \neq 0$.) 
The action in Eq.(\ref{S5:VA}) takes the same form as in 
Refs.~\cite{DaRold:2005zs,Erlich:2005qh,Hong:2006si,Piai:2006hy,Haba:2008nz}
except that $w(z) = L/z$ has been replaced by the one given in Eq.(\ref{w}).
Our induced metric determined by the equation of motion 
for the bulk scalar $\Phi_X$ is compared with the effective metric of
Ref.~\cite{Hirn:2005vk} 
lacking the bulk scalars $\Phi$ and $\Phi_X$, 
where the form of the effective metric was simply assumed to reproduce 
the OPE for the vector and axial-vector current correlators $\Pi_{V,A}$.

We solve the equations of motion for the transversely 
polarized components of the gauge fields $V_\mu(x,z)$ and $A_\mu(x,z)$. 
The corresponding equations of motion are immediately read off from the action (\ref{S5:VA}) as 
\begin{eqnarray}
&& 
\left[q^2+w(z)^{-1}\partial_z w(z)\partial_z\,\right]V_{\mu}(q,z)=0 
\,, \label{VEOM} \\
&& \left[q^2+w(z)^{-1}\partial_z w(z)\partial_z-
2\left(\frac{L}{z}\right)^{2} v^2(z)
\,\right]A_{\mu}(q,z)=0
\, ,
\label{AEOM}
\end{eqnarray}
where $V_\mu(q,z)$ and $A_\mu(q,z)$ denote 
the Fourier transforms of $V_\mu(x,z)$ and $A_\mu(x,z)$, respectively. 
It is convenient to decompose $V_\mu(q,z)$ and $A_\mu(q,z)$ 
into the external sources $(\tilde{v}_\mu(q), \tilde{a}_\mu(q))$ 
and the remainders $(V(q,z), A(q,z))$, such as  
$V_{\mu}(q,z) = \tilde{v}_{\mu}(q)\,V(q,z)$ and 
$A_{\mu}(q,z) = \tilde{a}_{\mu}(q)\,A(q,z) $. 
Using the equations of motion (\ref{VEOM}) and (\ref{AEOM}) 
together with the boundary conditions in Eq.(\ref{VAmuBC}), 
we rewrite the action (\ref{S5:VA}) to get  
the four-dimensional UV boundary term which is holographically 
dual to the generating functional $W[v_\mu, a_\mu]$ in TC,   
\begin{equation}
W[v_\mu,a_\mu] = \frac{1}{2}\int \frac{d ^4q}{(2\pi)^4} \,
\frac{-1}{g_5^2}w(\ep)
\tr \left[\tilde{v}_{\mu}(-q)\partial_zV(q,\ep)\cdot \tilde v^{\mu}(q)
+\tilde a_{\mu}(-q)\partial_zA(q,\ep)\cdot \tilde a^{\mu}(q)\right]
\,, 
\end{equation} 
where $V(q, z)$ and $A(q, z)$ respectively satisfy the boundary conditions, 
 \beqs
\label{VBCs}
V(q,\ep) &=& 1 \, , \qquad 
\partial_z V(q,\zm) = 0\,, \\
\label{ABC}
A(q,\ep) &=& 1\,,  \qquad 
\partial_z A(q,\zm) = 0 
\,. 
\eeqs

    From Eqs.(\ref{Pi:V}) and (\ref{Pi:A}), 
the vector and axial-vector current correlators 
$\Pi_{V}$ and $\Pi_A$ are now expressed in terms of 
the five-dimensional gauge theory as 
\begin{eqnarray} 
\Pi_V(Q^2) = \frac{w(\ep)}{g_5^2}\partial_zV(Q^2,\ep) 
 \,, \label{Pi:Vp} \qquad
\Pi_A(Q^2) = \frac{w(\ep)}{g_5^2}\partial_zA(Q^2,\ep) 
\,, \label{Pi:Ap}
\end{eqnarray}
where we have rewritten $V(q,z) = V(Q^2,z)$ and $A(q,z)=A(Q^2,z)$. 
We emphasize that,  thanks to the introduction of the bulk scalar field $\Phi_X$ dual to 
the role of the gluon condensation, 
the present model reproduces $\GB/Q^4$ term 
in the OPE of the current correlators $\Pi_V$ and $\Pi_A$,
which was missing in the previous approach 
without $\Phi_X$~\cite{DaRold:2005zs,Erlich:2005qh,Hong:2006si,Piai:2006hy,Haba:2008nz}.
Leaving the details of the derivation in Appendix~\ref{OPE},
we will here just show the resultant expression of the high-energy expansion 
of $\Pi_{V,A}(Q^2)$ in the large Euclidean-momentum region $(1/z_m)^2 \ll Q^2  < (1/\ep)^2$, 
\beqs
\Pi_{V,A}(Q^2) \Bigg|_{(1/z_m)^2 \ll Q^2 < (1/\ep)^2} 
= Q^2 \left[~\frac{L}{2 g_5^2}\ln{Q^2} 
+ c \frac{2}{3} \frac{g_5^2}{L} \frac{\langle \alpha G_{\mu\nu}^2 \rangle}{Q^4} 
+{\cal{O}}(\frac{1}{Q^{6-2\gamma_m}})
~\right]
\,. 
\label{PiV:complete0}
\eeqs 
Furthermore, by introducing 
some extra higher-dimensional interaction terms (See Eq.(\ref{S5:int})), 
the present model exactly 
reproduces the high-energy behavior up to terms suppressed 
by $(1/Q^8)$, consistently with the form of the OPE, 
\beqs
\Pi_{V,A}(Q^2) \Bigg|_{(1/z_m)^2 \ll Q^2 < (1/\ep)^2} 
= Q^2 \left[~\frac{L}{2 g_5^2}\ln{Q^2} 
+ c \frac{2}{3} \frac{g_5^2}{L} \frac{\langle \alpha G_{\mu\nu}^2 \rangle}{Q^4} 
+ C_6^{V,A} L^{2 \gamma_m} \frac{\condense^2_{1/L}}{Q^{2(3-\gamma_m)}}
+{\cal{O}}(\frac{1}{Q^8})
~\right]
\,,
\label{PiV:complete}
\eeqs 
where the couplings $C_6^{V,A}$ come from 
the higher-dimensional interaction terms (See Eq.(\ref{C6})).  
It should also be stressed that such extra interaction terms do not affect 
all of our results shown in the later sections.

Our model is sharply contrasted with the approach in
Ref.~\cite{Hirn:2005vk}
where the effective metric was assumed so as to produce $1/Q^4$ term 
which, however, could be confused with the chiral condensation term
$\condense^2/Q^{6-2\gamma_m} \simeq \condense^2/Q^4$ in the case of SWC-TC
with $\gamma_m \simeq 1$ in the absence of the bulk scalars $\Phi$ and $\Phi_X$.
In our case which explicitly includes the bulk scalar field $\Phi_X$ dual to the 
gluon condensate, we are able to obtain not just the form behaving as $(1/Q^4)$ 
but the whole expression $\langle \alpha  G_{\mu\nu}^2 \rangle/Q^4$ involving 
the gluon condensate $\langle \alpha G_{\mu\nu}^2 \rangle$,
and hence clearly distinguish from the 
$\langle \bar{T}T \rangle^2/Q^4$ term 
arising due to the bulk scalar $\Phi$.

In order to obtain the formulas for the decay constant $F_\pi$ and 
the $S$ parameter,  we shall expand $\Pi_{V,A}(Q^2)$ perturbatively in powers of 
 $Q^2$ as $ \Pi_{V,A}(Q^2) = \Pi_{V,A}(0)+Q^2\Pi'_{V,A}(0)+ {\cal{O}}(Q^4) $, 
where $ \Pi'_{V,A}(0)  \equiv \partial\Pi_{V,A}(Q^2)/ \partial Q^2 \Big|_{Q^2=0}$.
Then $\Pi_{V,A}(0)$ and $\Pi_{V,A}'(0)$ are expressed as  
\begin{eqnarray}
\Pi_V(0)  &=& 0 
\label{PiV0}\,, 
\hspace{110pt} 
\Pi_V'(0) 
=-\frac{L}{g_5^2}\int_{\ep}^{z_m}\frac{ d\,z^{\prime}}{z^{\prime}}
\chi^2(z')\,, \label{PiV'0} \\   
\Pi_A(0)  &=& \frac{L}{g^2_5 }\frac{1}{\ep}
\chi^2(\ep)\partial_z A(0,z) \Bigg|_{z=\ep}
\,, 
\label{PiA0} 
\qquad 
\Pi_A'(0)
=- \frac{L}{g_5^2}\int_{\ep}^{z_m}\frac{ d\,z^{\prime}}{z^{\prime}}
\chi^2(z') A^2(0,z')
\,, \label{PiA'0}  
\end{eqnarray}
where $A(0, z)$ is given as a solution to Eq.(\ref{AEOM}) with $q^2=0$. 
  From Eqs.(\ref{fpi:def}) and (\ref{Sdif}), we find that $F_\pi$ and $\hat{S}$ are expressed 
  in terms of the five-dimensional gauge theory as 
\begin{eqnarray} 
F_\pi^2 
&=& 
- \frac{L}{g^2_5 }\frac{1}{\ep}
\chi^2(\ep)\partial_z A(0,z) \Bigg|_{z=\ep} 
\,, \label{fpi:sol}
\\ 
\hat{S} &=& 
 4\pi\frac{L}{g_5^2}\int_{\ep}^{z_m}\frac{ d\,z}{z}
\chi^2(z) (1- A^2(0,z)) 
\,. \label{S:sol} 
\end{eqnarray}

\subsection{Vector, axial-vector, and flavor-singlet scalar meson
  masses: $M_{V_n}$, $M_{A_n}$, and $M_{\sigma_n}$}

A set of vector meson masses $\{M_{V_n}\}$ arises as  
an infinite tower of eigenvalues of normalizable solutions $\{V_n(z)\}$ 
satisfying Eq.(\ref{VEOM}) 
with $q^2 $ replaced by $\{M_{V_n}^2 \}$, 
\begin{eqnarray}
\left[M_{V_n}^2+w(z)^{-1}\partial_z w(z)\partial_z\,\right]V_n(z)=0 
\,, \label{VEOM:mass}
\end{eqnarray}
with the boundary condition 
$V_n(\ep) = 0$ and $\partial_z V_n(\zm) = 0$. 
The lowest eigenvalue is identified as the techni-$\rho$ meson mass, $M_{V_1}=M_\rho$.

 Similarly for axial-vector meson masses $\{M_{A_n}\}$, 
the eigenvalue equation for a set of normalizable modes $\{A_n(z) \}$ 
is obtained by taking $q^2=M_{A_n}^2$ in Eq.(\ref{AEOM}):  
\begin{equation}
\left[M_{A_n}^2+w(z)^{-1}\partial_z w(z)\partial_z
-2\left(\frac{L}{z}\right)^{2} v^2(z)\,\right]A_n(z)=0
\,, \label{AEOM:mass}
\end{equation} 
with the boundary condition $A_n(\ep)=0$ and $\partial_z A_n(z_m)=0$. 
The lowest eigenvalue is regarded as the techni-$a_1$ meson mass,
$M_{A_1}=M_{a_1}$.

The equation of motion for the flavor-singlet scalar field 
$\sigma(x,z)$ is decomposed into the eigenvalue equations 
for a set of the KK modes $\sigma^{(n)}(x)$ arising as 
$\sigma(x,z)=\sum_{n=1}^\infty \sigma^{(n)}(x) \sigma_n(z)$. 
By taking into account 
Eq.(\ref{Phi:decompose}) and replacing 
the momentum-squared $q^2$ with the mass-squared $M_{\sigma_n}^2$,  
the equation of motion for the wave function $\sigma_n(z)$ 
 is read off from the action (\ref{S5}) as
\begin{equation}\label{SnEOM}
\left[M_{\sigma_n}^2+\left(\frac{w(z)}{z^2}\right)^{-1}
\partial_z \left(\frac{w(z)}{z^2} \right) \partial_z
-\frac{(3-\gamma_m)(1+\gamma_m)}{z^2}
\,\right]\sigma_n(z)=0 
\, ,\label{SEOM:0}
\end{equation}  
where the normalizable solution $\sigma_n(z)$ should 
satisfy the UV boundary condition, 
$\lim_{\epsilon \to 0} \sigma_n(\epsilon)=0$,
so as to make the action finite at $z=\epsilon$ in the limit $\epsilon \to 0$.
The solution to Eq.(\ref{SEOM:0}) is then given as 
\begin{equation} 
\label{fS:sol}
\sigma_n(z)=c_\sigma \frac{z^2}{1+G \left(\frac{z}{z_m}\right)^4} 
J_{1-\gamma_m}(M_{\sigma n}z)
\,, 
\end{equation}
which satisfies the UV boundary condition. In Eq.(\ref{fS:sol}) 
$J_n$ is Bessel function of order $n$ 
and the overall coefficient $c_\sigma$ is determined by a normalization 
condition which is irrelevant to the present study.

The eigenvalue equation for $M_{\sigma_n}$ is provided by 
the IR boundary condition which needs a bit careful consideration 
because it is related to a vacuum stabilization problem: 
The vacuum energy on the background of the bulk scalar field $\Phi$ is 
equivalent to the generating functional in Eq.(\ref{W:MM'}). 
By using the solution for $v(z)$ in Eq.(\ref{v:sol}), the vacuum energy is 
expressed as a function of $\xi$ in the chiral limit $M=0$, 
 \beqs 
V[\xi]= - \int d^4 x \frac{L}{2g_5^2} \chi^2(z_m)
\tr\left[\left( 3 -\gamma_m -4\frac{G}{1+G}\right)\frac{\xi^2}{z_m^4}\right]
\,, 
\eeqs
where we have put $M'=0$ for simplicity. 
One can easily see that, with respect to $\xi$, 
$V[\xi]$ is minimized at $\xi=0$, which readily 
leads to $\langle \bar{T} T \rangle=0$ through Eq.(\ref{condense:sol}) 
 and hence to no spontaneous breaking of chiral symmetry. 
In order to avoid this problem, 
similarly to a procedure proposed in Ref.~\cite{DaRold:2005vr},  
we may introduce the following IR potential: 
\beqs
{\cal L}_{\rm IR} &=& -\left(\frac{L}{z}\right)^4
\chi^2(z) V(\Phi)\Bigg|_{\zm} ,\nonumber \\
 V(\Phi)|_{z=z_m}&=&
-m^2_b\Tr|\Phi(z_m)|^2+
\lambda\Tr(|\Phi(z_m)|^2)^2
\,,  
\label{pot}
\eeqs 
where the potential parameters $m_b^2$ and $\lambda$ are 
taken to be positive.
By adding this IR potential, the vacuum is now realized 
at $\xi \neq 0$:   
\beqs
\xi^2 = \frac{1}{\lambda}\left(L^2 m_{b}^2  
- \frac{L}{g_5^2} \left(3- \gamma_m -4 \frac{G}
{1 + G}\right)\right)
\,,
\label{mb}
\eeqs
where $\xi^2$ is tuned to be positive by adjusting $m_b^2$.
The IR boundary condition for $\sigma_n(z)$ is now assigned in a way 
that the total IR boundary term with respect to $\sigma^{(n)}(x)$ 
is canceled in the quadratic order:  
\beqs
\label{SIRBC}
\big[\partial_z +
2\left( \frac{L}{z_m}  \right) g_5^2 m^2_{\sigma 5}\big]\sigma_n(z) \Bigg|_{z_m}=0,
\eeqs
where
\beqs
m_{\sigma 5}^2 = \frac{1}{L^2} \left[ 
 \lambda \xi^2- \frac{1}{2} 
\frac{L}{g_5^2}
\left(3 -\gamma_m -4 \frac{G}
{1 + G}\right)
\right] 
\,. 
\label{ms}
\eeqs

Substituting the solution in Eq.(\ref{fS:sol}) into the IR boundary condition (\ref{SIRBC}),
we thus obtain the eigenvalue equation for $M_{\sigma_n}$,  
\beqs
2 \lambda \xi^2\frac{g_5^2}{L} J_{1 -\gamma_m}(M_{\sigma n}z_m)
=M_{\sigma n}z_m \cdot J_{2-\gamma_m}(M_{\sigma n}z_m) 
\,.
\label{SEOM}
\eeqs
The lowest eigenvalue is identified as 
the techni-dilaton mass, $M_{\sigma_1}= M_{\rm TD}$.

\section{Analysis on gluonic-contributions} 
\label{sec3}

In this section, we shall discuss effects on observables coming from 
the gluon condensation 
in a generic TC with $0 \lesssim \gamma_m \lesssim 1$ involving the case of QCD with $\gamma_m\simeq 0$. 
Among observables, we particularly focus on the $S$ parameter and 
the masses of the lowest KK modes for the vector, axial-vector, and flavor-singlet scalar mesons
$(M_\rho, M_{a_1}, M_{\rm TD})$. 
 To this end, we first pay our attention to the parameters  
describing the present five-dimensional model and 
momentarily discuss how they can be fixed by considering the holographic-dual of the generic TC 
with $0 \lesssim \gamma_m \lesssim 1$. 
The parameters are following ten:  
\begin{equation} 
\frac{L}{g_5^2}, \qquad 
z_m, \qquad 
\epsilon, \qquad
\gamma_m, \qquad 
\xi, \qquad 
M, \qquad   
G (\textrm{or ${\cal G}$}), \qquad 
M', \qquad 
c, \qquad 
\lambda 
\,. 
\end{equation} 

The UV brane position $\ep$ is treated as the cutoff scale ($1/\ep$) 
of the five-dimensional theory and is usually set to be $0$ after all calculations are done. 
From a point of view of a typical TC scenario, on the other hand, 
the UV cutoff $(1/\ep)$  
can be replaced by an ETC scale $(1/\ep) = \Lambda_{\rm{ETC}}$. 
 As for the IR brane position $z_m$, similarly, it can play a role of 
the IR cutoff scale of the theory associated with the chiral symmetry 
breaking or confinement, and hence ($1/z_m$) can be related to
 a typical meson mass scale, say, $m$, in TC.  
Since $(m/\Lambda_{\rm ETC})  \ll 1$, 
we may simply put $(\ep/z_m) = (m/\Lambda_{\rm ETC})=0$. 
Then we see that the $S$ parameter does not depend on either 
$\ep$ or $z_m$ since it is a dimensionless quantity. 
Note, however, that other dimensionful quantities, such as $F_\pi$, $M_\rho$, $M_{a_1}$, 
and $M_{\rm TD}$, still have a certain $z_m$-dependence which can be completely factorized by defining 
dimensionless ones like $\tilde{F}_\pi=z_m F_\pi$, and so on.

The parameter $M'$ is the external source of the techni-gluon 
condensation-operator $\alpha G_{\mu\nu}^2$ and hence is 
regarded as the explicit breaking source of the conformal/scale symmetry 
associated with the dilatation current anomaly characterized by the
intrinsic scale of order $\Lambda_{\rm TC}$ 
($\simeq \Lambda_{\rm ETC} \gg 1/z_m$ for $\gamma_m \simeq 1$)
in which we are not interested. 
 Here we take $M'=0$.
(Even when $M'=0$ and $\ep = 0$, we have the explicit breaking of conformal/scale invariance 
due to $1/z_m \neq 0$.)

The parameters ($L/g_5^2$) and $c$ are determined 
by comparing the high-energy behavior of the current correlators $\Pi_{V,A}$ 
to those obtained by the OPE: 
($L/g_5^2$) from the $\ln{Q^2}$ term and $c$ from the $\GB/Q^4$ term.
In the case of SWC-TC with $\gamma_m \simeq 1$, 
$\langle \bar{T}T \rangle^2/Q^4$ term has the same $Q^2$-dependence as
that of $\langle \alpha G_{\mu\nu}^2 \rangle/Q^4$ term.
As was discussed in Sec.~\ref{2B}, it is possible to clearly distinguish 
those two terms in our approach. 
We will leave the detailed calculation in Appendix~\ref{OPE} and here just quote 
the result on the OPE-matching: 
\begin{equation} 
\frac{L}{g_5^2} = \frac{N_{\rm TC}}{12 \pi^2}
\,, \qquad 
c = -\frac{N_{\rm TC}}{192 \pi^3} 
\,. \label{g5-c}
\end{equation}

 The parameter $\lambda$ has been introduced so as to minimize 
 the bulk scalar potential with non-zero $\xi$.  
 In order to know more about $\lambda$,  
let us take a look at the two terms in the square bracket of Eq.(\ref{ms}). 
One then finds that the first term ($\lambda \xi^2$) should be proportional to $N_{\rm TC}$ 
because of Eq.(\ref{g5-c}). 
  Furthermore,  from Eqs.(\ref{condense:sol}) and (\ref{g5-c}) 
and taking into account $\langle \bar{T} T \rangle \propto N_{\rm TC}$, 
we see that $\xi \propto N_{\rm TC}^0 $ and hence 
$\lambda \propto N_{\rm TC}$. 
Supposing that  
the coupling $\lambda$ 
is expected to be generated at one-loop level 
(through techni-fermion loops), 
we may totally write $\lambda=\kappa \times N_{\rm {TC}}/(4\pi)^2$ 
with an ${\cal O}(1)$ parameter $\kappa$: 
\begin{equation} 
 \lambda=\kappa \,\frac{N_{\rm {TC}}}{(4\pi)^2},
\qquad
\kappa=1.0 \,(\pm 0.3)
\,, \label{kappa}
\end{equation} 
which reproduces the mass of  
the flavor-singlet scalar bound-state 
(two-quark state), $f_0(1370)$, in QCD
as the lightest KK mode 
of the flavor-singlet scalar, $M_{\sigma_1}$.
Actually in QCD, 
there are other two candidates 
for the light flavor-singlet scalar particles 
other than $f_0(1370)$, which are $f_0(600)$ (so-called $\sigma$) and $f_0(980)$. 
The following is the reason why we have adopted $f_0(1370)$ 
to fit the value of $M_{\sigma_1}$: 
Since we are interested in application to a generic TC,  
we need to carefully select a certain appropriate flavor-singlet scalar bound-state  
realized in a generic strongly coupled dynamics 
with arbitrary $N_c$ and $N_f$. Thinking about the other 
two candidates from this point of view, one finds that 
those bound-states can be considered as four-quark states 
due to a specific property arising only in the case of 
real life QCD with $N_f=N_c=3$ (See, for example, Ref.~\cite{Hooft:2008we}), 
so they are excluded from candidate of 
two-quark state. 
We further notice that 
similar characteristic features in real life QCD would cause 
mixing between two-quark states and four-quark states, 
which could make the observed mass of the two-quark state $f_0(1370)$ 
lifted up. Without such a mixing, the mass of $f_0(1370)$ 
is expected to be around $1.1$-$1.2$ GeV~\cite{Black:1999yz}
\footnote{
Actually in Ref.\cite{Black:1999yz},  
the mixing between $a_0(980)$ and
$a_0(1445)$ is discussed. 
It is known, however, that in the large $N_c$ limit  
$f_0(1370)$ and $a_0(1445)$ are degenerate, 
so we can apply the similar argument against the mixing between $f_0(980)$ and $f_0(1370)$.  
}.
Thus we have determined the values of $\kappa$ as in Eq.(\ref{kappa}) ($0.7\le  \kappa \le 1.3$) 
using a set of QCD-fit values (See Eq(\ref{QCD-para-fit})) so as to reproduce the allowed range of 
the mass of $f_0(1370)$ without the mixing.

Thus we are now left with the four undetermined parameters,
\beq 
\label{4para}
z_m,\qquad \xi,\qquad G, \qquad \gamma_m\, (0 \lesssim \gamma_m \lesssim 1), 
\eeq
with $z_m$ being the only dimensionful parameter.

\subsection{QCD case with $\gamma_m\simeq 0$}

We shall first consider the case of QCD with $N_c=3$ and $\gamma_m\simeq 0$.  
In the case of QCD, 
the parameters $\xi$, $G$, and $(1/z_m)$ can be fixed 
in such a way that Eqs.(\ref{fpi:sol}), (\ref{VEOM:mass}), and (\ref{GB:sol}) 
respectively reproduce the experimental values~\cite{Amsler:2008zzb}
$f_\pi \simeq 92.4 \,\MeV$, 
$M_\rho \simeq 775 \, \MeV$, 
and
 a typical empirical value~\cite{Shifman:1978bx}
$\frac{1}{\pi} \GB  \simeq 0.012 \,\GeV^4$: 
\beqs
\xi \simeq 3.1\,, \qquad 
G \simeq  0.25\,,  \qquad 
\frac{1}{z_m} \simeq 347 \, {\rm MeV}
\,. \label{QCD-para-fit}
\eeqs  
Using these values, we calculate 
the values of $M_{a_1}$ (from Eq.(\ref{AEOM:mass})), 
$M_{\sigma_1}\equiv M_{f_0(1370)}$ (from Eq.(\ref{SEOM}) and using 
the values of $\kappa$ in Eq.(\ref{kappa})), 
$\hat{S}\,(=-16\pi L_{10})$ (from Eq.(\ref{S:sol})), and the quark condensate $\langle \bar{q} q \rangle$ 
(from Eq.(\ref{condense:sol})), which results in good agreement with experiment  
as shown in Table.~\ref{table:QCD}. 

\begin{table}
\begin{center}
\begin{tabular}{|c|c|c|c|c|} 
\hline 

QCD-fit 
 & $M_{a_1}$ & $M_{\sigma_1=f_0(1370)}$ 
& $\hat{S} = -16\pi L_{10}
$ & $(-\VEV{\bar qq})^{1/3}$\\
\hline \hline 
this model & 1264 MeV 
& $1.15 \,(^{+0.04}_{-0.07})\,\GeV$
& $0.31$ & 277 $\MeV$ \\ 
\hline 
measured & 1230 $\pm$ 40 MeV~\cite{Amsler:2008zzb}
&*1.1-1.2 GeV & $0.33 \pm 0.04$~\cite{Harada:2003jx} 
& 225 $\pm$ 25 $\MeV$~\cite{Gasser:1982ap} \\
\hline    
\end{tabular} 
\caption{ 
Values of several observables in QCD predicted from the present holographic model. 
The starred measured value of $M_{\sigma_1 =f_0(1370)}$ 
corresponds to the mass estimated without 
mixing with a four quark state $f_0(980)$ (See discussion below
 Eq.(\ref{kappa})).}
\label{table:QCD}
\end{center}
\end{table}

In our model, meson masses have some
contributions from the gluon condensation. 
The effect of the gluon condensation on $M_\rho$ can 
be analytically estimated for $G\simeq 0.25$
by expanding Eq.(\ref{VEOM:mass}) perturbatively in $G$:  
\begin{eqnarray}
\label{QCD:fit}
M_\rho &\simeq&  
\frac{2.41}{z_m} \times 
[1 - 0.388 \, G ] \nonumber \\ 
&\simeq & 
836 \, {\rm MeV} \times \left[ (1.000)_{G=0} - (0.097)_{G\neq 0} \right] \simeq 775 \, {\rm MeV} 
\,, \label{Mrho:G-dep:val}
\end{eqnarray} 
where in reaching the last line we have used $(1/z_m)\simeq 347$ MeV in
Eq.(\ref{QCD-para-fit}). 
Equation (\ref{Mrho:G-dep:val}) implies that the gluon condensation
negatively contributes 
about $10\%$ to the $\rho$ meson mass in QCD. 
Similar expression was obtained 
in somewhat different approach~\cite{Hirn:2005vk}.

\subsection{Generic TC case with $0 \lesssim \gamma_m \lesssim 1$} 
\label{TCcase}

We next discuss the case of a generic TC with $0 \lesssim \gamma_m \lesssim 1$ 
involving SWC-TC with $\gamma_m \simeq 1$~\footnote{
One might think that walking/near conformal dynamics is characterized by not $\gamma_m=1$ 
but $\gamma_m\simeq 1$, as in a typical example of SWC-TC based 
on the Caswell-Banks-Zaks infrared fixed
point~\cite{Caswell:1974gg,Banks:1981nn}  in the large $N_f$ QCD.  
However, as was clarified in Ref.~\cite{Haba:2008nz},  
in the present holographic approach, 
there exists no discontinuity between $\gamma_m=1$ 
and the limit $\gamma_m \to 1$, so that both cases give the same result. 
In the present work, therefore, we have explicitly taken $\gamma_m=1$. }  
and evaluate gluonic-contributions to 
the $S$ parameter $(S=\hat{S}\cdot(N_f/2))$, masses of techni-$\rho$ ($M_\rho$), 
techni-$a_1$ ($M_{a_1}$), and the lightest flavor-singlet scalar,
techni-dilaton ($M_{\rm TD}= M_{\sigma_1}$).

As we noted in Eq.(\ref{4para}), 
the dimensionless quantity $\hat{S}$ in Eq.(\ref{S:sol}) is given as a function of 
the three dimensionless parameters $\xi$, $\gamma_m$, and $G$,  
\begin{equation} 
\hat{S} = \hat{S} \left(\gamma_m, \xi, G  \right)  
\,.  \label{para:S} 
\end{equation}
Other dimensionful quantities, $F_\pi$, $M_\rho$, $M_{a_1}$, and $M_{\rm TD} $ 
are respectively expressed as follows: 
\begin{eqnarray}  
F_\pi &=& z_m^{-1} \cdot \widetilde{F_\pi} \left(\gamma_m, \xi, G\right)  
\,, \label{para:fpi} 
\\
M_{\rho} &=& z_m^{-1} \cdot \widetilde{M_\rho} \left(G \right)  
\,, \label{para:mrho} \\
M_{a_1} &=& z_m^{-1} \cdot \widetilde{M_{a_1}} \left(\gamma_m, \xi, G \right)  
\,, \label{para:a1} \\
M_{\rm TD} &=& z_m^{-1} \cdot \widetilde{M_{\rm TD}} 
\left(\gamma_m, \xi; \kappa \right)      
\,. \label{para:ms} 
\end{eqnarray} 
As in QCD, we will take the value of $\kappa$ 
in Eq.(\ref{kappa}), $\kappa = 1.0\, (\pm 0.3)$.
Absence of explicit dependence of $\xi$ and $\gamma_m$ in $M_\rho$ 
can be seen in Eq.(\ref{VEOM:mass}).
Explicit $\xi$- and $\gamma_m$-dependences for $M_{a_1}$ enter in the $v^2(z)$ term in 
Eq.(\ref{AEOM:mass}).
$M_{\rm TD}$ has no explicit $G$-dependence as seen in Eq.(\ref{SEOM}).
Note that every dimensionful quantities scales with the parameter $(1/z_m)$. 
Hereafter we shall consider dimensionless quantities 
$(M_\rho/F_\pi),  (M_{a_1}/F_\pi)$,  and $(M_{\rm TD}/F_\pi)$ 
which are free from $z_m$.

To properly evaluate effects from 
the gluon condensate $\GB$ in Eq.(\ref{GB:sol}),  
we need to be a bit careful because the expression (\ref{GB:sol}) 
involves not only $G$, the source of the gluon condensation, but also 
the IR cutoff scale $(1/z_m)$. 
This implies that a naive variation of the value of $\GB$ would not merely lead to  
the change of the value of the gluon condensation itself. 
In order to extract the gluonic-contribution only, therefore, 
we shall use $\GB/F_\pi^4$ which is not including $(1/z_m)$.  
It is convenient, furthermore, to work on the quantity normalized by the QCD value: 
\beqs
\label{B:def}
\Gamma = \Gamma(\gamma_m,\xi,G) \equiv \left(\frac{\left(\frac{1}{\pi}\GB/F_\pi^4\right)}
{\left(\frac{1}{\pi}\GB/f_\pi^4\right)_{\rm QCD}}\right)^{1/4}
= \left(\frac{\left(\frac{1}{\pi}\GB/F_\pi^4\right)}
{\left(0.012\,\GeV^4/(92.4\,\MeV)^4\right)_{\rm QCD}}\right)^{1/4}
\,.
\eeqs

In Fig.~\ref{mras:B:S3} 
we show the plots of ($M_\rho/F_\pi$), ($M_{a_1}/F_\pi$), and $(M_{\rm TD}/F_\pi)$ 
as a function of $\Gamma$ for $\gamma_m \simeq 0$ (left panel) 
and $\gamma_m \simeq 1$ (right panel) with $N_{\rm TC} = 3$ 
and $\hat S= 0.31$ (QCD value) fixed, 
where in calculating $(M_{\rm TD}/F_\pi)$ we have used $\kappa = 1.0$. 
 It is interesting to note that, 
for both cases with $\gamma_m \simeq 0$ and $\gamma_m \simeq 1$,  
{\it $(M_\rho/F_\pi)$ and $(M_{a_1}/F_\pi)$ tend to coincide to be degenerate constant ($\simeq 8.0$) 
as $\Gamma$ increases,  fairly independent of the value of the $S$ parameter}: Such a  degenerate spectrum of the techni-$\rho$ and techni-$a_1$
relatively independent of
the $S$ parameter is a salient feature of the
large contribution of  the techni-gluon condensate and would have  characteristic impact on the techni-hadron phenomenology. 
On the other hand, the {\it techni-dilaton mass  $(M_{\rm TD}/F_\pi)$ is very sensitive to $\Gamma$,
rapidly decreasing as $\Gamma$ increases.}
We have checked that these behaviors do not alter even for different values of ${\hat S}$ 
other than ${\hat S}=0.31$.

 \begin{figure}
\begin{center}
\includegraphics[width=6cm]{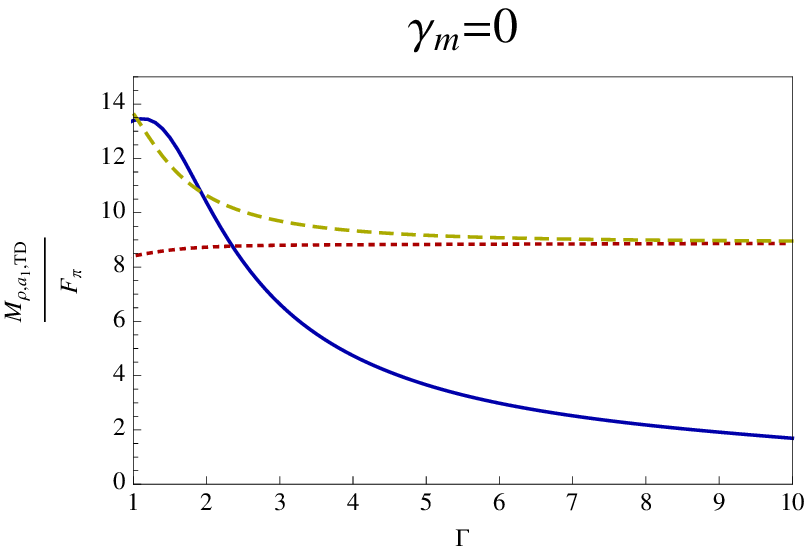}
\includegraphics[width=6cm]{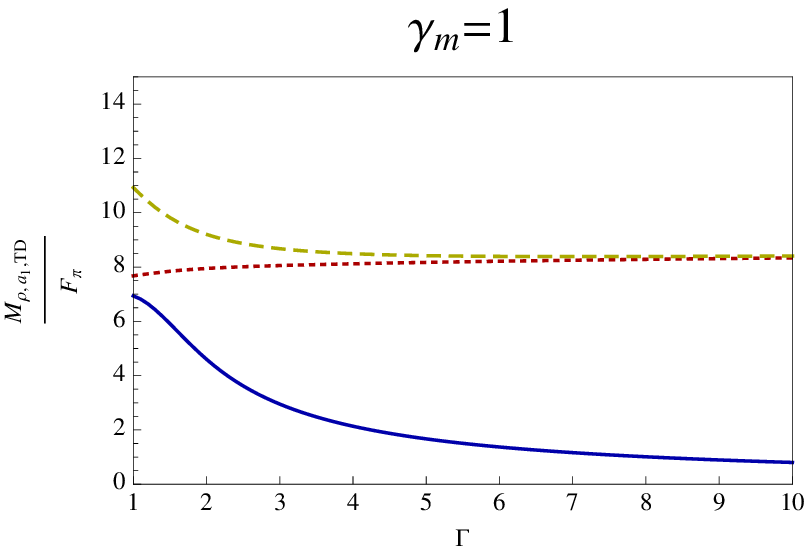}
 \caption{\label{mras:B:S3}
Plots of $(M_{\rho}/F_\pi)$, $(M_{a_1}/F_\pi)$, and $(M_{\rm TD}/F_\pi)$
as a function of $\Gamma$ with $N_{\rm TC} = 3$ and $\hat S=0.31$ fixed. 
The left panel and the right panel respectively correspond to 
the cases with $\gamma_m \simeq 0$ and $\gamma_m = 1$. 
In both panels, 
the dotted (red), dashed (yellow), and solid (blue) curves 
respectively denote $(M_{\rho}/F_\pi)$, $(M_{a_1}/F_\pi )$,  
and $(M_{\rm TD}/F_\pi)$. 
In the left panel, the values on the vertical axis ($\Gamma = 1$)
correspond to those of QCD.
} 
\end{center} 
 \end{figure}

It is also worth studying the $\gamma_m$-dependences of 
$(M_\rho/F_\pi)$, $(M_{a_1}/F_\pi)$, and $(M_{\rm TD}/F_\pi)$.
Figure~\ref{mras:ad} shows the plots of $(M_\rho/F_\pi)$,
$(M_{a_1}/F_\pi)$, and $(M_{\rm TD}/F_\pi)$
as a function of $\gamma_m$ for $\hat S= 0.31$ (left panel) and  
$\hat S = 0.1$ (right panel) with $N_{\rm TC} = 3$ and $\Gamma=1$ fixed. 
  Looking at Fig.~\ref{mras:ad}, we see that 
$(M_\rho/F_\pi)$ and $(M_{a_1}/F_\pi)$ are not sensitive to  
$\gamma_m$, while $(M_{\rm TD}/F_\pi)$ rapidly decreases 
as $\gamma_m$ becomes larger
for both cases with $\hat S = 0.31$ and $\hat S = 0.1$. 
Actually, in the case of  $\hat S \le 0.1$, 
such a relatively large suppression on $(M_{\rm TD}/F_\pi)$ generated by the large anomalous dimension 
can be seen by analytically solving Eq.(\ref{SEOM}) for $(M_{\rm TD} z_m) \ll 1$~\footnote{  
The condition $(M_{\rm TD} z_m) \ll 1$ is satisfied only when $\hat S < 0.1$. } 
expanding Eq.(\ref{SEOM}) in power of $(M_{\rm TD} z_m)$: 
\beqs 
\label{MS:ana}
(M_{\rm TD} z_m)^2 \simeq (2-\gamma_m) \cdot 4 \lambda \xi^2  \frac{g_5^2}{L}
=   (2-\gamma_m) \cdot 3\kappa \xi^2,  
\eeqs
where we have used Eqs.(\ref{g5-c}) and (\ref{kappa}).

 \begin{figure}
\begin{center}
 \includegraphics[width=6cm]{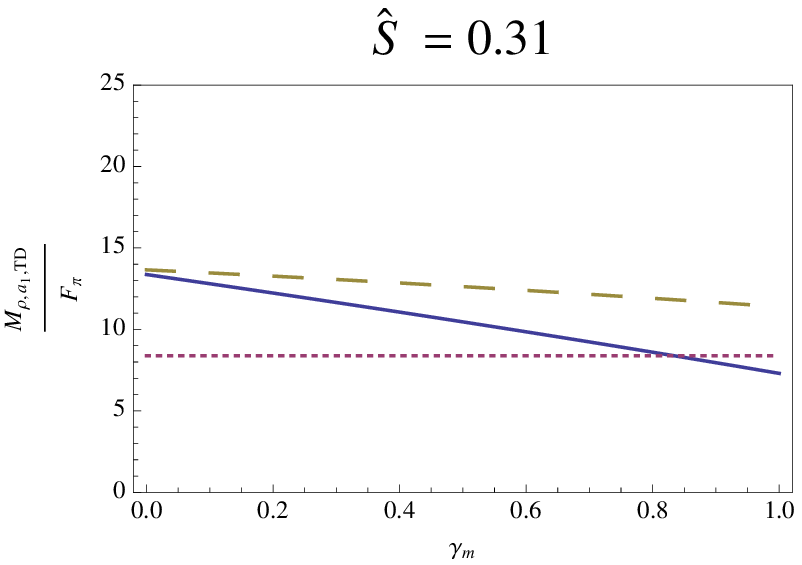} 
\qquad \includegraphics[width=6cm]{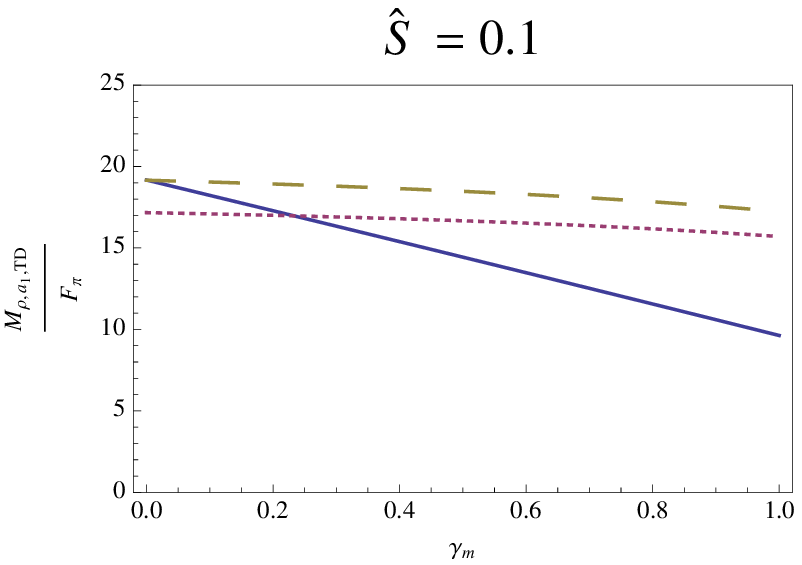}
 \caption{\label{mras:ad}
Plots of $(M_{\rho}/F_\pi)$, $(M_{a_1}/F_\pi)$, and $(M_{\rm TD}/F_\pi)$ 
as a function of $\gamma_m$ with $N_{\rm TC}=3$ and $\Gamma=1$ fixed. 
The left panel corresponds to the case with $\hat S=0.31$ while the right panel $\hat S = 0.1$. 
The dotted (red), dashed (yellow), solid (blue) lines 
respectively denote $(M_{\rho}/F_\pi)$, $(M_{a_1}/F_\pi )$,  
and $(M_{\rm TD}/F_\pi)$ in both panels. 
} 
\end{center}
\end{figure}

Let us next focus on the case of  a generic SWC-TC with $\gamma_m \simeq 1$ 
and evaluate contributions from the techni-gluon condensation 
to the $S$ parameter $(S={\hat S}\cdot (N_f/2))$.
In Fig.~\ref{S:Bfm:comp} we show the plot of $\hat{S}$ divided by 
$N_{\rm TC}$ as a function of $(F_\pi/M_\rho)^2$ for $\Gamma= 0,5,10$. 
In this figure ($\hat{S}/N_{\rm TC}$) is
restricted in a phenomenologically arrowed range, 
$(\hat{S}/N_{\rm TC}) \lesssim 0.05$~\footnote{The value of the upper bound, 0.05,  
can be estimated as follows: Consider a conservative upper bound of the $S$ parameter, 
$S \lesssim 0.1$ and look at the relationship with ${\hat S}$, 
$(S/N_{\rm TC})=(N_f/2)\cdot (\hat{S}/N_{\rm TC})$. 
Since $N_{\rm TC}\ge 2$ and $N_f \ge 2$ (i.e. the number of techni-doublets $N_{\rm TD} \ge 1$), 
one then finds that $S \lesssim 0.1$ leads to 
$(\hat{S}/N_{\rm TC}) \lesssim (0.1/[N_{\rm TC}]_{\rm min}) \times (2/[N_f]_{\rm min})=0.05$. }. 
Looking at Fig.~\ref{S:Bfm:comp}, we can easily see that $\hat S$ 
monotonically decreases  with respect to $(F_\pi/M_\rho)^2$ 
in both cases with $\Gamma=0$ and $\Gamma \neq 0$ and 
is continuously approaching zero as $(F_\pi/M_\rho)^2 \to 0$.
Conversely, when we tune $\hat{S}$ smaller, 
($M_{\rho}/F_\pi$) gets larger (($M_{a_1}/F_\pi$) and
($M_{\rm TD}/F_\pi$) as well):
\beq 
\label{mf:S}
(M_{\rho}/F_\pi,\, M_{a_1}/F_\pi,\,M_{\rm TD}/F_\pi) \nearrow\,\,\,\, {\rm as}
\,\,\,\,\hat S\searrow.
\eeq
This tendency coincides with what the authors found in Ref.~\cite{Haba:2008nz}. 
Indeed, the model with $\Gamma=0$ 
is nothing but the one analyzed in Ref.~\cite{Haba:2008nz}. 
Note that for $ \Gamma = 5$ and $10$ the techni-gluon condensation reduces 
the $S$ parameter up till 10\% in the phenomenologically acceptable 
region of $S$ ($S = \hat{S}\cdot (N_f/2) \lesssim 0.1$).

 \begin{figure}
\begin{center}
 \includegraphics[width=6cm]{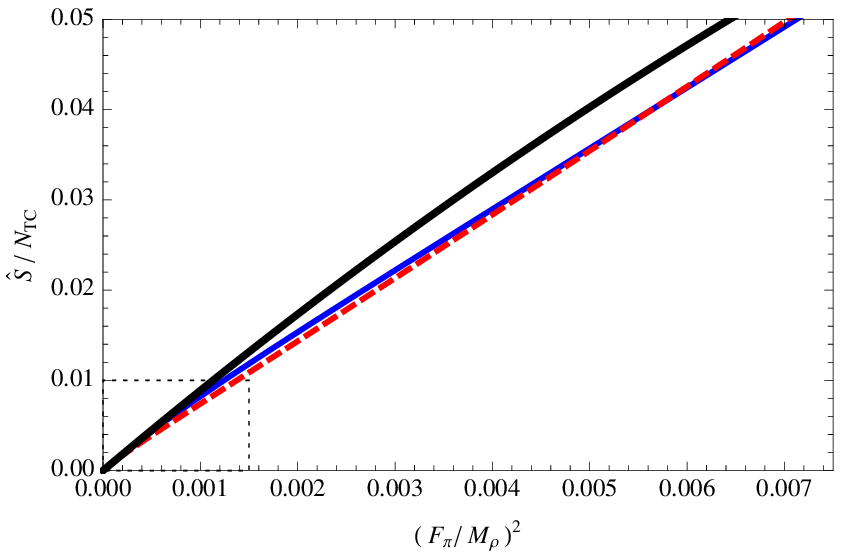}
 \includegraphics[width=6cm]{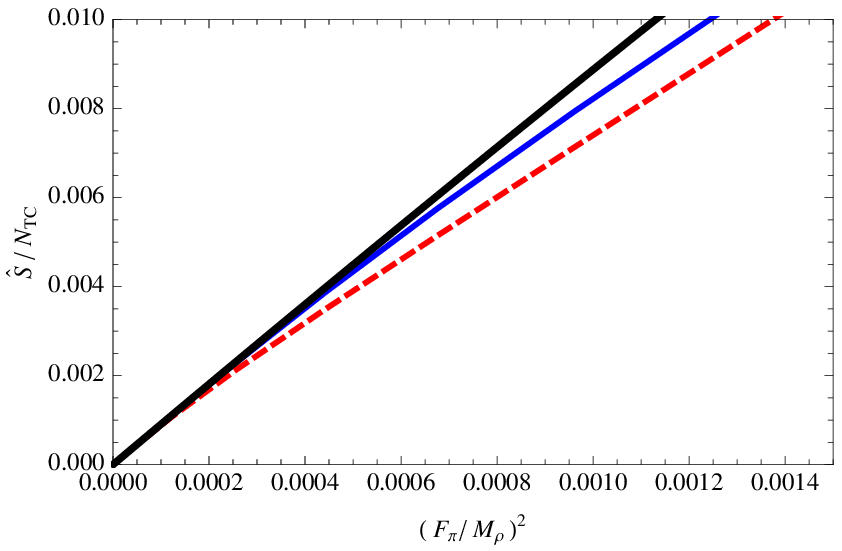}
 \caption{\label{S:Bfm:comp}
Plots of $\hat S/N_{\rm TC}$ as a function of $\left(F_\pi/M_\rho\right)^2$ 
in the case of a generic SWC-TC with $\gamma_m\simeq 1$. 
In the left panel ${\hat S}/N_{\rm TC}$ 
is restricted in a phenomenologically interesting 
region ${\hat S}/N_{\rm TC} \lesssim 0.05$. 
The right panel focuses on the dashed-rectangle area (${\hat S}/N_{\rm TC} \lesssim 0.01$) 
in the left panel. 
In both panels, the thick (black) line  represents the case with $\Gamma=0$
which corresponds to the previous analysis without $G$~\cite{Haba:2008nz}, 
while the thin (blue) line and the dashed (red) line respectively correspond  
to the cases with $\Gamma=5$ and $\Gamma=10$.  } 
\end{center} 
 \end{figure}

To see a general tendency of $(M_{\rho}/F_\pi)$, $(M_{a_1}/F_\pi)$,  
and $(M_{\rm TD}/F_\pi)$ with respect to $\Gamma$, we shall 
fix $\hat S$ to be a certain value,  
say, $\hat{S} =0.1$ (minimal requirement for a realistic TC)~\footnote{
Actually, the phenomenological bound for the $S$ parameter is 
$S \lesssim 0.1$, not $\hat{S}(=S/(N_f/2)) \lesssim 0.1$. 
Both constraints would coincide only when $N_f=2$ (minimal flavors). 
More detailed discussion including the flavor-dependence 
will be given in the next section. }. 
 For ${\hat S} = 0.1$ and $N_{\rm TC} = 3$, 
we evaluate the $\Gamma$-dependences of $(M_\rho/F_\pi)$,
$(M_{a_1}/F_\pi)$, and $(M_{\rm TD}/F_\pi)$  
to get the plots shown in Fig.~\ref{mras:B}. 
Here we have taken $\kappa=1$  in calculating $M_{\rm TD}$ 
and have chosen $\Gamma =10$ as the maximal value 
which turns out to be large enough 
when we consider typical models of SWC-TC as will be seen later.  
Figure~\ref{mras:B} tells us that, for $0 \le \Gamma \le 10$, 
the input $\hat S=0.1$ constrains the meson masses to be respectively in the following ranges:   
$15.7  \ge (M_\rho/F_\pi) \ge 14.6,\,
17.3 \ge (M_{a_1}/F_\pi) \ge 14.7 ,\,
9.79 \ge (M_{\rm TD}/F_\pi) \ge 1.45 $. 
This implies that the gluonic-contribution decreases 
$(M_\rho/F_\pi)$, $(M_{a_1}/F_\pi)$, and $(M_{\rm TD}/F_\pi)$ 
about 9\%, 15\%, and 85\%, respectively, 
during the value of $\Gamma$ evolves from 0 to 10. 
Note that, although we have not specified any types of SWC-TC yet,  it is remarkable that each
value of the $M_\rho$ and $M_{a_1}$ is fairly insensitive to $\Gamma$ but  becomes a degenerate non-zero constant  as $\Gamma$ increases,
the value being almost determined just by fixing the $S$ parameter. (The degeneracy itself is realized no matter what value the $S$ might take.)
 In contrast, the mass of techni-dilaton $M_{\rm TD}$ is very sensitive
to $\Gamma$, rapidly decreasing as $\Gamma$ increases. 
In fact  $M_{\rm TD}/F_\pi \to 0$ in the formal  limit $\Gamma \to \infty$. 
This tendency is still operative even if $S$ is much smaller,
although the decreasing rate of $(M_{\rm TD}/F_\pi)$ becomes somewhat milder.   
In the next section we will discuss a matching with a couple of concrete models of SWC-TC 
in which the value of $\Gamma$ is related to 
$(\Lambda_{\rm ETC}/F_\pi)$ in such a way that $\Gamma \simeq 6$ -- $8$
for $(\Lambda_{\rm ETC}/F_\pi) = 10^4$ -- $10^5$.   
 ($\Gamma \to \infty$ for $\Lambda_{\rm ETC}/F_\pi \to \infty$.)

 \begin{figure}
\begin{center}
 \includegraphics[width=8cm]{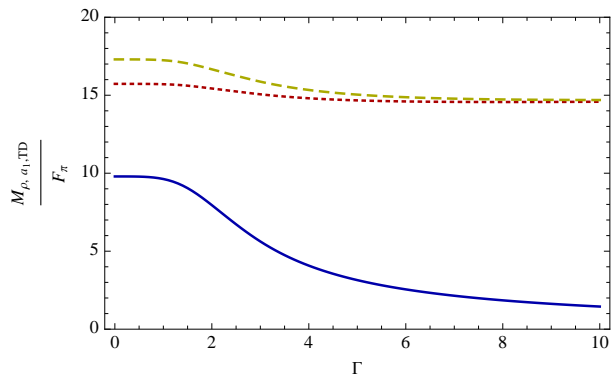}
\caption{\label{mras:B}
Plots of $(M_{\rho}/F_\pi)$, $(M_{a_1}/F_\pi )$, and $(M_{\rm TD}/F_\pi)$
as a function of $\Gamma$ for $\hat S = 0.1$ 
in a generic SWC-TC with $N_{\rm TC}=3$ and $\gamma_m \simeq 1$. 
The dotted (red), dashed (yellow), and solid (blue) curves respectively correspond 
to $(M_{\rho}/F_\pi)$, $(M_{a_1}/F_\pi)$, and $(M_{\rm TD}/F_\pi)$.  
}  
\end{center} 
 \end{figure}

\section{Matching with typical models of SWC-TC based on large $N_f$ QCD} 
\label{ladder:match}

In this section, we consider 
typical models of SWC-TC based on the Caswell-Banks-Zaks IR fixed
point~\cite{Caswell:1974gg,Banks:1981nn} 
$\alpha_*$ of the two-loop beta function
in the large $N_f$ QCD, QCD with massless flavors $3 \ll N_f<11N_{\rm TC}/2 $. 
It has been suggested that there exists a certain region 
$(N_f^* <) N_f^{\rm cr} < N_f < 11N_{\rm TC}/2$ 
(which is called ``conformal window'') 
such that $\alpha_* < \alpha_{\rm cr}$, 
where the critical coupling $\alpha_{\rm cr}$ for the spontaneous
chiral symmetry breaking
may be estimated as $\alpha_{\rm cr}=\pi/(3C_2(F))$ 
based on the Schwinger-Dyson (SD) equation in the ladder approximation.
Equating  $\alpha_* = \alpha_{\rm cr}=\pi/(3C_2(F))$, 
we find $N_f^{\rm cr}\simeq 4 N_{\rm TC}$~\cite{Appelquist:1996dq}
for the conformal phase transition point. 
There are many studies about the existence of the conformal window and 
the value of $N_f^{\rm cr}$ (if exists) 
in various non-perturbative methods 
including lattice gauge theories~\cite{SCGT09}.
Here we simply assume the existence of the conformal window 
and tentatively use the value of $N_f^{\rm cr}$ from the
perturbative two-loop beta function and ladder SD equation.

It was argued~\cite{Miransky:1996pd} that the conformal phase transition is characterized by 
scaling of the essential singularity 
(Miransky scaling~\cite{Miransky:1984ef})
\beqs
\label{mdyn}
m \simeq 4 \Lambda 
\exp{\left(-\frac{\pi}{\sqrt{\alpha/\alpha_{\rm cr}-1}}\right)}
\,, 
\eeqs
where $m$ is the dynamical mass of techni-fermion and 
$\Lambda$ an intrinsic scale ($\Lambda_{\rm TC}$)
which may be identified with 
an ETC scale $\Lambda_{\rm ETC}$ 
in the actual model building:
\beq
\Lambda = \Lambda_{\rm TC} \simeq \Lambda_{\rm ETC} \gg m.
\eeq
We can arrange a large hierarchy $m \ll \Lambda$ 
in terms of (approximately) conformal symmetry
by tuning the theory close to the conformal fixed point 
as $\alpha \simeq \alpha_* \to \alpha_{\rm cr}$ 
in the broken phase $\alpha_*> \alpha_{\rm cr}$,
in such a way that the coupling constant is almost non-running
over the wide range $m<\mu < \Lambda$ (see Fig.~\ref{fig:beta}).
(There still exists a remnant of the conformal symmetry 
due to the IR fixed point $\alpha_*$, 
although the IR fixed point $\alpha_*$ actually disappears
because techni-fermions with mass $m$ 
decouple from the beta function for $\mu < m$.)

In the SWC dynamics near the conformal window, 
the {\it explicit} breaking of the conformal symmetry 
manifest as the conformal anomaly 
is due to the generation of the dynamical mass of techni-fermions $m$
which arises from the {\it spontaneous} breaking of the conformal and chiral symmetry.
Thus the techni-gluon condensation for the conformal anomaly 
relevant to the dynamics near the conformal window 
is directly related to $m$ 
but not the intrinsic scale $\Lambda_{\rm TC}\gg m$.
The conformal anomaly for this dynamical generation 
takes the following form~\cite{Miransky:1989qc}: 
\beqs
\label{pD}
\VEV{\partial^\mu D_\mu} 
= \VEV{\theta^{\mu}_{\mu}} 
=4 \VEV{ \theta_{0}^{0}}
=\lim_{\Lambda \to \infty}
\frac{\beta_{\rm NP}(\alpha)}{4 \alpha^2} \GB, 
\eeqs
where $D_\mu$ and $\theta_{\mu}^\nu$ are the dilatation current and 
the energy-momentum tensor, respectively, and $\beta_{\rm NP}(\alpha)$ denotes 
the {\it non-perturbative} beta function of the gauge coupling $\alpha$ 
related to the generation of $m$ in Eq.(\ref{mdyn}):
\beqs
\beta_{\rm NP}(\alpha)  
\equiv  \frac{\partial \alpha}{\partial \ln{\Lambda}} 
= -\frac{2}{3 C_2(F)}
\left(\frac{\alpha}{\alpha_{\rm cr}}-1\right)^{\frac{3}{2}}
= -\frac{2\pi^3}{3 C_2(F)}
\left(\ln{4\frac{\Lambda}{m}}\right)^{-3} 
\,, \label{beta}
\eeqs 
in which we have used $\alpha_{\rm cr}=\pi/(3C_2(F))$. 
Straightforward calculation of the effective potential 
at two-loop order~\cite{Miransky:1989qc} yields 
the vacuum energy $\VEV{\theta_0^0}$:  
\beqs
\label{theta00}
\VEV{\theta_0^0} = -\frac{N_f N_{\rm TC}}{\pi^4}m^4.
\eeqs 
From Eqs.(\ref{pD}) and (\ref{theta00}), 
we obtain
\begin{equation} 
 \GB 
=   - \frac{16}{\pi^4} \frac{\alpha^2}{\beta_{\rm NP}(\alpha)} N_f N_{\rm TC}\,m^4 
\,.  \label{G2:SD}
\end{equation}

Note that Eq.(\ref{pD}) takes the same form as 
a conventional conformal anomaly obtained 
in the all order perturbation theory~\cite{Adler:1976zt},
where we naturally expect 
$\GB = {\cal{O}}(\Lambda_{\rm TC}^4) \gg {\cal{O}}(m^4)$.
Here we ignore techni-glueball dilaton with mass 
of this order $\Lambda_{\rm TC}$ 
associated with the perturbative anomaly and 
the running effect of the coupling 
for $\mu > \Lambda_{\rm TC}$.
In the ordinary QCD with $N_c = N_f = 3$,
we have $\GB = {\cal{O}}(\Lambda_{\rm QCD}^4) = {\cal{O}}(m^4)$:
There is no such a large hierarchy $m \ll \Lambda_{\rm QCD}$
and conformal region $m < \mu < \Lambda_{\rm QCD}$ 
where the coupling constant is almost non-running 
$\alpha(Q) \simeq {\rm{ constant }}$.
In contrast, our techni-gluon condensate $\GB$ in Eq.(\ref{pD}) 
is responsible for the conformal anomaly 
induced by the dynamical generation of mass $m$
and hence $\GB = {\cal{O}}(m^4) \ll {\cal{O}}(\Lambda_{\rm TC}^4)$
and our techni-dilaton is a bound state of techni-fermions
with mass $m$ which breaks the conformal symmetry near the conformal window. 
Also note that the conformal region 
with the almost non-running coupling is realized only
when we arrange $N_f$ as $N_f \propto N_{\rm TC}$ in such a way that 
the fermionic dynamics and the gluonic dynamics cooperate intimately. 
Thus in Eq.(\ref{G2:SD}) we have 
$\VEV{G_{\mu \nu}^2} \sim {\cal{O}}(N_f N_{\rm TC}) \sim {\cal{O}}(N_{\rm TC}^2)$ 
in accord with large $N_{\rm TC}$ counting 
relevant to holographic models.

From Eqs.(\ref{mdyn}), (\ref{beta}), and (\ref{G2:SD}), 
the techni-gluon condensate $\GB$ 
is expressed in terms of $m$ and $\Lambda$ as 
\beqs
\label{GB:lad}
\GB|_{\alpha \simeq \alpha_{*}} 
&=& 
\lim_{\Lambda \to \infty}
\frac{8}{3} \frac{N_f N_{\rm TC}}{ C_2(F) \pi^2} 
m^4
\left(\frac{\ln{\frac{4\Lambda}{m}}}{\pi}\right)^{3}
\left(1+
\left(\frac{\ln{\frac{4\Lambda}{m}}}{\pi}\right)^{-2}\right)^2 
\,. 
\eeqs
Comparing this to Eq.(\ref{GB:sol}) 
with Eq.(\ref{g5-c}) taken into account, 
we arrive at a relationship between $G$, $(z_mm)$, and $(\Lambda_{\rm ETC}/m)$,  
\beqs
G
=
C\cdot 
(z_m m )^4 
\left(\frac{\ln{\frac{4\Lambda_{\rm ETC}}{m}}}{\pi}\right)^{3}
\left(1+
\left(\frac{\ln{\frac{4\Lambda_{\rm ETC}}{m}}}{\pi}\right)^{-2}\right)^2
\,, 
\label{GB:match}
\eeqs
where 
\beq
C =\frac{1}{2\pi} \frac{N_f N_{\rm TC}}{N_{\rm TC}^2-1} 
\,. 
\eeq

In terms of $m$, the techni-fermion condensate $\langle \bar{T} T \rangle_m$ 
renormalized at $\mu=m$ can be evaluated as
\footnote{
Numerically $t$ coincides
with the prefactor in Eq.(\ref{condense:sol}), 
$3- \gamma_m$, for $\gamma_m \simeq 0,\,1,\,2$.
In the case of QCD with $\gamma_m \simeq 0$, 
this implies $m \simeq 453\, \MeV$ for 
the value of $\VEV{\bar qq}\simeq -(277 \, \MeV)^3$
in accord with the conventional
constituent quark mass $m  \simeq 350\, \MeV$
and with $m \simeq 420 \, \GeV$ from the Pagels-Stokar (PS) formula.
For details see Appendix.~\ref{PS:cond}
}
\begin{equation} 
 \condense_m = Z_m \cdot\condense_{\Lambda} =  - t \cdot \frac{N_{TC}}{4\pi^2}  m^3 
 \,\qquad {\rm with}\,\,t \simeq 2,
 \label{scaling-law}
\end{equation}
where the mass renormalization constant is $Z_m = Z_m(m/\Lambda) = m/\Lambda$.
We regard the condensate  in Eq.(\ref{condense:sol}) 
as the one renormalized at $\mu = (1/L)$ following the procedure suggested in Ref.~\cite{Haba:2008nz}. 
We then find the parameter $\xi$ is related to $G$ together with $(z_m m)$ as follows: 
\beqs
\xi 
= \frac{\sqrt{3}}{1+G}(z_m m)^{2}
\,, 
\label{xi:match}
\eeqs 
where we have used Eq.(\ref{g5-c}). 
   From Eqs.(\ref{xi:match}) and (\ref{GB:match}), we see that the 
two parameters $\xi$ and $G$ are now replaced 
by $(z_m m)$ and $(\Lambda_{\rm ETC}/m)$ involving the quantities concerning the SWC-TC, 
the dynamical mass $m$ and the ETC scale 
$\Lambda_{\rm ETC}$.

To be concrete for our analysis, 
we will assume the value of $N_f^{\rm cr}$ 
as that from the two-loop beta function 
and ladder SD equation $N_f^{\rm cr} \simeq 4 N_{\rm TC}$
bearing in mind the large $N_{\rm TC}$ limit in accord with holographic setup.

\begin{table}

\begin{center} 
\begin{tabular}{|c|c|c|c|c|}
\hline 
\hspace{1pt} $N_{TC}$ \hspace{1pt}  & 
\hspace{1pt} $\log_{10}(\Lambda_{\rm {ETC}}/F_\pi)$ \hspace{1pt}  & 
\hspace{20pt} $\xi$ \hspace{20pt} 
&\hspace{20pt}  $G$ \hspace{20pt}  
&\hspace{2pt}  $1/z_m$ [TeV] \hspace{2pt} 
\\ 
\hline \hline  
2&4 &  0.22 & 1.12 & 2.09 
\\\hline 
2&5 & 0.15 & 1.69 & 2.31 
\\\hline \hline  
3&4 &  0.21 & 0.60 &1.90 
\\\hline 
3&5 & 0.16 & 0.97 & 2.04 
\\\hline \hline  
4&4 &  0.19 & 0.37 & 1.82 
\\\hline 
4&5 & 0.15 & 0.62 & 1.92 
\\\hline 
\end{tabular} 
\caption{Values of the model-parameters 
fitted to the SWC-TC with $N_{\rm TC}= 2,3,4$ 
and $N_f=N_f^{\rm EW} = 4 N_{\rm TC}$ based on 
the large $N_f$ QCD. Here use has been made of 
$S={\hat S} (N_f^{\rm EW}/2) = 0.1$. 
}  
\label{table:para}
\end{center} 

\end{table}

  Let us now recall Eqs.(\ref{para:S})-(\ref{para:ms}) which 
imply that all the quantities given  in those equations (say, $S=S(\xi, G)$ and $F_\pi=z_m^{-1} \tilde{f}_\pi(\xi, G)$)   
are determined once we fix the one dimensionful parameter $z_m$ and 
the two dimensionless parameters $(\xi,G)$ which are now rephrased 
by $(z_m m)$ and $(\Lambda_{\rm ETC}/m)$.  
To fix $\xi$ or $G$, 
we may use a certain value of the $S$ parameter (e.g. $S=0.1$) 
as a phenomenological input in such a way as was done in the previous section. 
To determine the size of $z_m$, 
we can use the familiar formula 
\beq 
F_\pi=246/\sqrt{N_f^{\rm EW}/2} \, \GeV 
\,, 
\eeq
where $N_f^{\rm EW}$ is the number of techni-fermions 
belonging to the doublet of $SU(2)_L$ symmetry in the standard model (SM) 
and may be different from the number of techni-fermion flavors $N_f$
which participate in the SWC-TC dynamics.  
Furthermore, the ETC scale $\Lambda_{\rm ETC}$ 
 may be constrained to be in a range, 
$10^3 \lesssim \Lambda_{\rm ETC} \lesssim 10^4$ TeV:
\beq
\label{lamda:range}
10^4 \lesssim \Lambda_{\rm ETC}/F_\pi \lesssim 10^5 
\,, 
\eeq 
so as to accommodate the realistic light quark masses without suffering from 
the flavor changing neutral current (FCNC) syndrome. 
Use of these inputs now fixes the values of the parameters 
$(\xi, G, 1/z_m)$. 
In Table~\ref{table:para} we list these values 
for each $N_{\rm TC}=2,3,4$, 
and $(\Lambda_{\rm ETC}/F_\pi)=10^4,10^5$ 
in the case of $N_f = N_f^{\rm EW} = 4 N_{\rm TC}$ 
and $S = \hat S (N_f^{\rm EW}/2) = 0.1$.

\begin{table}
\begin{center}
\begin{tabular}{|c|c|c|c|c|c|c|c|c|} 
\hline 

\hspace{1pt} $N_{TC}$ \hspace{1pt}  
& \hspace{1pt} $\log_{10} (\Lambda_{\rm ETC}/F_\pi)$ \hspace{1pt} 
&  \hspace{10pt} $M_{\rm TD}$ [GeV]  \hspace{10pt} 
&  \hspace{10pt} $M_\rho$ [TeV]  \hspace{10pt} 
&  \hspace{10pt} $M_{a_1}$ [TeV]  \hspace{10pt} 
&\hspace{20pt}  $\Gamma$ \hspace{20pt}
&\hspace{2pt}  $m$ [TeV] \hspace{2pt} 
&\hspace{2pt} $m/m_{\rm PS}$ \hspace{2pt}  
&\hspace{2pt} $R$ \hspace{2pt}  

\\ 
\hline \hline 
2&4 & 777 $\left(^{+106}_{-125}\right)$ &3.75& 3.82&5.93 &1.08 &1.49&1.11 \\ 
\hline 
2&5 & 613 $\left(^{+85}_{-99}\right)$ &3.69 & 3.74&7.26&1.13& 1.57&1.16\\ 
\hline \hline 
 3&4 & 681 $\left(^{+94}_{-110}\right)$ &3.86 & 3.90&6.26&0.84& 1.74 & 1.48\\ 
\hline 
3&5 &  556 $\left(^{+77}_{-90}\right)$ &3.80 & 3.83& 7.57&0.87&1.80&1.53\\ 
\hline \hline 
4&4 & 597  $\left(^{+82}_{-96}\right)$ &3.93 & 3.95&6.58&0.71&1.97 &1.84\\ 
\hline 
4&5 & 505  $\left(^{+70}_{-82}\right)$ &3.87 &3.89 &7.88 &0.73&2.02&1.88\\ 
\hline    
\end{tabular} 
\caption{ 
Estimates of $M_\rho$, $M_{a_1}$, and 
$M_{\rm TD}$ for $S=0.1$ in the SWC-TC with 
$N_{\rm TC}= 2,3,4$ and $N_f=N_f^{\rm EW}=4 N_{\rm TC}$ based on 
the large $N_f$ QCD. 
The range of the values of $M_{\rm TD}$ come from 
varying the value of $\kappa$ in the range $0.7 \le \kappa \le 1.3$, 
where the smallest values of $M_{\rm TD}$ 
correspond to the cases with $\kappa=0.7$ 
while the largest values $\kappa=1.3$. 
$m_{\rm PS}$ and $R$ are defined in the text.} 
\label{table:mass1}
\end{center} 

\end{table}

Using the values of the parameters given in Table~\ref{table:para}, 
we calculate the masses of the 
techni-$\rho$ ($M_\rho$),  techni-$a_1$ ($M_{a_1}$), and techni-scalar ($M_{\rm TD}$) mesons 
to obtain Table~\ref{table:mass1}. 
The values of $M_{\rm TD}$ are estimated by varying the value of $\kappa$
from $0.7$ to $1.3$ (around 1.0 with 30\% error) 
in Table~\ref{table:mass1}. 
Note that $M_\rho$ and $M_{a_1}$ 
are almost degenerate to be 
$M_\rho \simeq M_{a_1} \simeq$ $3.7$--$3.9$ TeV 
for every $N_{\rm TC} = 2,3,4$,  in accord with the general tendency of the model with large techni-gluon condensate in section \ref{sec3} where  
the degeneracy was not linked to the smallness of the $S$ parameter and would have a new phenomenological
implications. 
In contrast, the techni-dilaton mass $M_{\rm TD} \sim 500$--$800\,\GeV$ (when $\kappa=1$) is much lighter than 
$M_\rho$ and $M_{a_1}$ also
in accord with the generic analysis for large $\Gamma$ in the previous section, 
although $\hat S = S/(N_f^{\rm EW}/2) = 0.1/(2N_{\rm TC})$ 
is somewhat smaller than $\hat S = 0.1$ used in the generic analysis 
in Fig.~\ref{mras:B}. 
Indeed, in the present case we have $\Gamma  \simeq  6$--$8$ 
as is seen from Table~\ref{table:mass1}. 
The essential reason for the large $\Gamma$ 
is due to the existence of 
the wide conformal region $F_\pi < \mu < \Lambda_{\rm ETC}$
with $(\Lambda_{\rm ETC}/F_\pi) = 10^4$--$10^5$,
which yields the smallness of the beta function 
through the factor $(\ln{4\Lambda_{\rm ETC}/m})^{-3}$
in Eq.({\ref{beta}) and hence amplifies 
the techni-gluon condensation in Eq.(\ref{G2:SD})
compared with the ordinary QCD with $\Gamma=1$. 
Note that  in the idealized (phenomenologically uninteresting) limit $\Lambda_{\rm ETC}/F_\pi \to \infty$ 
we would have $\Gamma \to \infty$ and hence $M_{\rm TD}/F_\pi \to 0$. 
(This does not mean that techni-dilaton becomes a true (exactly massless) NG boson, since 
its decay constant diverges, $F_{\rm TD}/F_\pi  \to \infty$,  in such an idealized limit and hence the techni-dilaton gets decoupled. See later discussions.)

Thus we would expect the techni-dilaton as a 
composite Higgs boson near the conformality of SWC-TC
with mass
\beq
M_{\rm TD} \simeq 600\, \GeV,
\eeq
while $M_{\rho}$ and $M_{a_1}$ are generally heavy:
\beq
M_\rho \simeq M_{a_1} \simeq 3.8\, \TeV.
\eeq

The values of the ratio $(m/m_{\rm PS})$ are also listed in Table.~\ref{table:mass1}, 
where $m_{\rm PS}$ denotes the dynamical mass estimated 
based on the PS formula: 
\beq
F_\pi^2 \simeq \frac{N_{TC}}{4\pi^2} m^2_{\rm PS} 
\cdot I,\qquad {\rm{with}}\,\,
I = \int_{0}^{\infty} d x x \frac{\Sigma^2(x)-\frac{x}{4}\frac{d}{d
x}\Sigma^2(x)}
{(x+\Sigma^2(x))^2} 
\,, \label{m:PS}
\eeq 
in which $\Sigma(x)\equiv \Sigma(Q^2)/m$ with $\Sigma(Q^2)$ being the mass function. 
When we use a simple-minded parametrization 
for $\Sigma(x)$, 
$\Sigma(x) = x^{(\gamma_m - 1)/2}$ for $x=Q^2/m^2 > 1$ and $\Sigma(x) = 1$ for $x< 1$, 
we get $I \simeq 1$ for SWC-TC with $\gamma_m \simeq 1$ 
while $I \simeq 0.6$ for QCD with $\gamma_m \simeq 0$ (For details see Appendix.~\ref{PS:cond}). 
It is interesting to note that $m/m_{\rm PS} > 1$. 
This can be understood by considering 
a PS formula appropriate for the present holographic analysis~\cite{Haba:2008nz}: 
\begin{equation}  
F_\pi^2 \simeq \frac{N_{\rm TC}}{4\pi^2} 
m^2 \cdot 
(z_m m)^{4 - 2\gamma_m} 
\,, \label{m:HPS}
\end{equation} 
which is satisfied for $(1/z_m) > m$ and $0\lesssim \gamma_m \lesssim 1$.  
  Using Eqs.(\ref{m:PS}) and (\ref{m:HPS}), certainly we find that 
$(m/m_{\rm PS}) \simeq (\frac{1/z_m}{m}) > 1$ for $\gamma_m\simeq1$.

Also listed is the ratio $R$ defined as 
\begin{equation} 
R^3=\sqrt{\frac{N_{\rm TC}}{3}}\frac{\condense_{m}/F_\pi^3}
{(\VEV{\bar qq}/f_\pi^3)_{\rm QCD}}
\,, 
\end{equation}  
which is slightly larger than 1
reflected by the result that $m/m_{\rm PS} > 1$, 
while it would be smaller than 1 ($R \simeq 0.69$) if the PS formula were used. 
Note that the enhancement of $R$ has nothing to do with 
that of $\condense_{\Lambda}$ caused by $Z_m^{-1}(\Lambda/m)$.

Let us now compare the present holographic model having the explicit  
techni-gluon contribution with the previous model~\cite{Haba:2008nz} 
without the techni gluon condensation. 
Were it not for the matching 
with the ladder SD analysis (namely without using Eqs.(\ref{mdyn}),(\ref{G2:SD}), and
(\ref{scaling-law})), 
the model of Ref.~\cite{Haba:2008nz} would be simply 
the $\Gamma = 0$ limit of the present model. 
Actually, the values of the meson masses lying on the $\Gamma=0$ line 
in Fig.~\ref{mras:B} are the results for $\hat S = 0.1$ in the previous model of Ref.~\cite{Haba:2008nz}. 
Here we compare the two models for the same value of $S$ as in Table.~\ref{table:mass1}. 
In Table~\ref{table:mass3} we show the values of the meson masses 
for $S=0.1$ and $N_{\rm TC}=2,3,4$ 
in the holographic SWC-TC without techni-gluon condensation~\cite{Haba:2008nz}, 
where we have used $\kappa =1.0$ in estimating the values of $M_{\rm TD}$. 
 Comparing the values in Tables~\ref{table:mass1} 
(present model with $\Gamma\simeq 6$--$8$) and~\ref{table:mass3} 
(previous model),  one hardly sees  
differences in $M_\rho$ and $M_{a_1}$, in accord with the general analysis in 
the previous section (see Fig.~\ref{mras:B:S3}), while 
there is a substantial difference in $M_{\rm TD}$ 
arising from the non-zero techni-gluon condensation as seen in Fig.~\ref{mras:B:S3}. 
Note that the relatively smaller $M_{\rm TD}$ 
compared with $M_\rho$ and $M_{a_1}$ is due to the large anomalous
dimension as was mentioned in Eq.(\ref{MS:ana}) and also seen from Fig.~\ref{mras:ad}.

\begin{table}[h]
\begin{center}
\begin{tabular}{|c|c|c|c|} 
\hline 

\hspace{10pt} $N_{\rm TC}$ \hspace{10pt}  
&  \hspace{20pt} $M_{\rm TD}$ [TeV]  \hspace{20pt} 
&  \hspace{20pt} $M_\rho$ [TeV]  \hspace{20pt} 
&  \hspace{20pt} $M_{a_1}$ [TeV]  \hspace{20pt} \\ 
\hline \hline 
2 & 1.58 &3.99& 4.13\\ 
\hline  
 3 & 1.08 &4.03 & 4.09\\ 
 \hline 
4 & 0.81 &4.04&4.08 \\ 
\hline    
\end{tabular} 
\caption{ 
Estimates of the meson mass in the holographic SWC-TC 
without techni-gluon condensation~\cite{Haba:2008nz} for $S=0.1$ 
with $N_{\rm TC} = 2,3,4$ and $N_f=4 N_{\rm TC}$. 
In calculating the values of $M_{\rm TD}$ 
the value of $\kappa =1.0$ has been used. } 
\label{table:mass3}
\end{center}
\end{table}

To summarize the characteristic feature of 
our holographic model,
we may compare our result for 
$(M_\rho/F_\pi,\, M_{a_1}/F_\pi,\,M_{\rm TD}/F_\pi)$
of the SWC-TC in Table~\ref{table:mass1}
with that of the ordinary QCD ($N_c = N_f = 3 $)
in Table~\ref{table:QCD}.
See Table.\ref{table:Mfpi} where 
we show the case of $N_{\rm TC}=3$ and $(\Lambda_{\rm ETC}/F_\pi)=10^4$ 
as a representative case of our SWC-TC model (the first row)
and the ordinary QCD with the input value of $(M_\rho/f_\pi)$
(the second row),
together with the $S=0.1$ case of the previous holographic SWC-TC model
without techni-gluon condensation ($\Gamma=0$)~\cite{Haba:2008nz} (the third row). 
Table~\ref{table:Mfpi} shows that 
the values of $(M_\rho/F_\pi)$ and ($M_{a_1}/F_\pi$) 
tend to become larger as the value of $S$ gets smaller
as noted in the generic analysis (See Eq.(\ref{mf:S})).
This tendency is actually almost independent of the values of $\Gamma$ and $\gamma_m$ 
as seen in Figs.~\ref{mras:B:S3} and~\ref{mras:ad}.   
Notably, although $(M_{\rm TD}/F_\pi)$ has the same tendency 
with respect to $S$, it receives other important effects: 
One is from the large anomalous dimension $\gamma_m \simeq1$ as 
in Fig.~\ref{mras:ad} and Eq.(\ref{MS:ana}),  
while the other from the large techni-gluon condensation $\Gamma=7.57$ 
as one can see from Figs.~\ref{mras:B:S3} and~\ref{mras:B}.

\begin{table}[h]
\begin{center}
\begin{tabular}{|l|c|c|c|} 
\hline 
\hspace{10pt} Holographic models with $N_{\rm TC}({\rm or}\,N_c) = 3$ \hspace{10pt}  
&  \hspace{20pt} $M_\rho/F_\pi$  \hspace{20pt} 
&  \hspace{20pt} $M_{a_1}/F_\pi$  \hspace{20pt} 
&  \hspace{20pt} $M_{\rm TD}/F_\pi$ \hspace{20pt} \\ 
\hline \hline 
SWC-TC with $\gamma_m \simeq 1$ ($\Gamma = 7.57, S=0.1$, $\Lambda_{\rm
 ETC}/F_\pi = 10^5$) 
& 37.8 &38.1 & 5.5\\ 
 \hline 
QCD with $\gamma_m\simeq 0$ ($ \Gamma=1, S=0.31$) & 8.4 &13.7& 12.4\\ 
\hline  
SWC-TC with $\gamma_m \simeq 1$ ($\Gamma=0, S=0.1$) &40.1 &40.7&10.7\\ 
\hline    
\end{tabular} 
\caption{ 
Comparison of $(M_\rho/F_\pi,\, M_{a_1}/F_\pi,\,M_{\rm TD}/F_\pi)$
between several holographic models with $N_{\rm TC}({\rm or}\, N_c)=3$. 
In calculating the values of $(M_{\rm TD}/F_\pi)$ 
$\kappa=1$ has been used. }
\label{table:Mfpi}
\end{center}
\end{table}

It is also worth comparing with the result of the straightforward 
calculation of the large $N_f$ QCD with $N_{\rm TC}=3$
based on the Bethe-Salpeter (BS) equation combined with the SD equation 
in the ladder approximation~\cite{Harada:2005ru,Harada:2003dc,Kurachi:2006ej}. 
The straightforward calculation of current correlators 
using the SD and BS equations has shown~\cite{Harada:2005ru} that $\hat{S} \simeq 0.17$~\footnote{
In Ref.~\cite{Harada:2005ru} 
the straightforward calculation of current correlators in the large $N_f$ QCD  
actually gives ${\hat S}\simeq 0.25$ for $N_{\rm TC}=3$. 
However, since it is known that the ladder SD and BS method has a tendency to 
overestimate ${\hat S}$ in QCD, which could be understood as scale ambiguity, 
the actual value near the conformal phase transition point with $\gamma_m \simeq 1$ 
has to be re-scaled by a factor about (2/3) so as to fit the QCD value properly 
in extending to the case of QCD. Then we obtain the re-scaled value of $\hat S$, 
$\hat{S}^{\rm re-scaled} \simeq 0.17$. } 
near the conformal phase transition point, 
for $\alpha_{*}/\alpha_{\rm cr} \simeq 1.13$, 
or equivalently, $(\Lambda_{\rm ETC}/F_\pi)=(\Lambda/F_\pi) \simeq 10^{4.6}$. 
To make a direct comparison, 
we take the same value of ${\hat S}$ and  
$(\Lambda_{\rm ETC}/F_\pi)=(\Lambda/F_\pi) \simeq 10^{4.6}$
($\Gamma \simeq 5.09$). 
The comparison of the meson masses normalized by $F_\pi$ 
is given in Table.~\ref{table:Mfpilad}. 
Here we have also shown the values with those in 
the previous holographic SWC-TC with $N_{\rm TC}=3$ 
without gluonic-condensation ($\Gamma =0$)~\cite{Haba:2008nz}. 
Looking at Table~\ref{table:Mfpilad}, one can see a curious 
coincidence among the values of $(M_\rho/F_\pi)$ and $(M_{a_1}/F_\pi)$ from different cases. 
This agreement does not depend so much on whether gluonic-contribution 
is turned off or not, as expected from Fig.~\ref{mras:B:S3}.  
On the other hand, $(M_{\rm TD}/F_\pi)$ is somewhat smaller than 
the value calculated from the ladder BS equation with the SD equation.

\begin{table}[h]
\begin{center}
\begin{tabular}{|l|c|c|c|} 
\hline 
\hspace{10pt} SWC-TC models with $\gamma_m \simeq 1$, $N_{\rm TC}=3$ and $\hat{S}=0.17$~\cite{Harada:2005ru} \hspace{10pt}  
&  \hspace{20pt} $M_\rho/F_\pi$  \hspace{20pt} 
&  \hspace{20pt} $M_{a_1}/F_\pi$  \hspace{20pt} 
&  \hspace{20pt} $M_{\rm TD}/F_\pi$ \hspace{20pt} \\ 
\hline \hline 
Holographic SWC-TC with $\Gamma=5.09$ ($\Lambda_{\rm ETC}/F_\pi =
 10^{4.6}$) 
&11.2 &11.5&2.3\\
\hline
Ladder BS with SD~\cite{Harada:2003dc,Kurachi:2006ej} &11.0 &11.5&4.2\\
\hline    
Holographic SWC-TC with $\Gamma=0 $ &11.6 &13.9 &8.9 \\ 
\hline

\end{tabular} 
\caption{ 
Comparison of the values of the meson masses normalized by $F_\pi$ 
between three SWC-TC models with $\gamma_m\simeq 1$ and $N_{\rm TC}=3$. 
In calculating the values of $(M_{\rm TD}/F_\pi)$ in the holographic models 
$\kappa=1$ has been used. $F_\pi$ in Ref.~\cite{Harada:2003dc,Kurachi:2006ej} 
is the value from the PS formula.}
\label{table:Mfpilad}
\end{center}
\end{table}

So far our analysis has been done for a specific value of $S$, $S =0.1$. 
As we already noted in Eq.(\ref{mf:S}), 
$(M_\rho/F_\pi,\,M_{a_1}/F_\pi,\,M_{\rm TD}/F_\pi)$ increase as $S$
decreases. 
Here we show how our result changes for different values of $S$ ($S \leq 0.1$).
In Fig.~\ref{S:mrhoms:B} we show the values of 
$M_{\rho}$ and $M_{\rm TD}$ for $\kappa=1$ in the SWC-TC with $N_{\rm TC}=3$ 
($N_f=4N_{\rm TC}=12$). 
The shaded area is the region obtained 
for $10^4 \le (\Lambda_{\rm ETC}/F_\pi)  \le 10^5$,  
so as to satisfy the phenomenological bound from the FCNC and $S \le 0.1$. 
Figure~\ref{S:mrhoms:B} indicates that, even 
when $0.07\lesssim S \leq 0.1$, 
$M_{\rm TD}$ can still lie in a region 
less than 1 TeV  
which is still in the discovery region at the Large Hadron Collider (LHC).

\begin{figure}
\begin{center}
 \includegraphics[width=8cm]{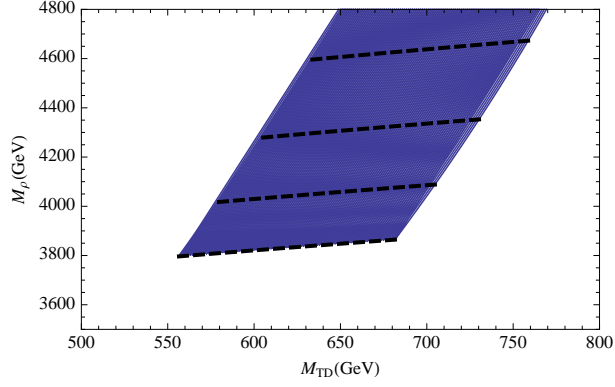}
 \caption{\label{S:mrhoms:B} 
Phenomenological constraints on masses of 
the techni-$\rho$ and techni-dilaton 
in the SWC-TC model with $N_{\rm TC}=3$ ($N_f=4N_{\rm TC}=12)$ 
based on the large $N_f$ QCD. 
The shaded region is drawn 
by varying the value of $(\Lambda_{\rm ETC}/F_\pi)$ 
from $10^4$ to $10^5$ so as to satisfy the phenomenological constraint $S\le 0.1$ 
and the FCNC. 
The dashed lines correspond to $S = 0.1,0.09,0.08,0.07$ from bottom to top. 
In the plot the value of $\kappa$ is taken to be 1. } 
\end{center} 
 \end{figure}

Finally, we calculate the decay constant of techni-dilaton, $F_{\rm TD}$,  
following a hypothesis of the partially conserved dilatation 
current (PCDC)~\footnote{
We could calculate $F_{\rm TD}$, not invoking the PCDC hypothesis,  
straightforwardly from the scalar-current correlator based on the holographic principle. 
Such an alternative (but direct) calculation would be pursued in future publications.} 
together with using the formula of the conformal anomaly given in Eq.(\ref{pD}). 
In this framework, the decay constant $F_{\rm TD}$ is given by~\cite{Bando:1986bg} 
\beqs
\label{fs}
F_{\rm TD}^2 = -4 \frac{\VEV{\theta_{\mu}^{\mu}}}{M_{\rm TD}^2} 
= \frac{16N_f N_{\rm TC}}{\pi^4}\frac{m^4}{M_{\rm TD}^2}
\,, 
\eeqs
where we have used Eq.(\ref{theta00}). 
This implies that  $F_{\rm TD}/F_\pi \sim  m/M_{\rm TD} \to \infty $ (decoupled techni-dilaton !) 
in the  idealized (phenomenologically uninteresting) limit $\Gamma\to \infty$ ($\Lambda_{\rm ETC}/F_\pi \to \infty$) 
where the techni-dilaton would be a true (exactly massless) Nambu-Goldstone boson, 
$M_\rho/F_\pi \sim M_{\rm TD}/m \to 0$. In our framework the conformal symmetry is 
always broken explicitly as well as spontaneously. 
Table~\ref{table:fs} shows a list of the values of $F_{\rm TD}$ 
for each $N_{\rm TC} = 2,3,4$ 
with $N_f= 4N_{\rm TC}$, $(\Lambda_{\rm ETC}/F_\pi) = 10^4$, and ${\hat S}=0.1$ fixed.  
The value of $F_{\rm TD}$ may be compared 
with that 
obtained based on the linear $\sigma$ model (LSM) approach 
which gives a formula between $F_{\rm TD}$ and $F_\pi$ as 
$F_{\rm TD}^{\rm LSM} = (3-\gamma_m) \sqrt{N_f/2} F_\pi =
(3-\gamma_m)\cdot  246 \, \GeV = 2 \times 246 \, \GeV$ 
(the third column of Table~\ref{table:fs}). 
The discrepancy between $F_{\rm TD}$ and $F_{\rm TD}^{\rm LSM}$
is not surprising since the SWC-TC dynamics is 
near the conformal phase transition where 
the Ginzburg-Landau/Gell-Mann-Levy effective theory (linear $\sigma$ model) 
breaks down~\cite{Miransky:1996pd}.

 We can estimate the Yukawa coupling constant of techni-dilaton 
by $F_{\rm TD}$ as 
\beq
g_Y^{\rm TD} = \sqrt{2} (3-\gamma_m) m_f/F_{\rm TD}
\,,
\eeq 
which is smaller than the Yukawa coupling in the SM, $g_{Y}^{\rm SM} 
= \sqrt{2} m_f/(246 \, \GeV) = \sqrt{2} (3-\gamma_m) m_f/F_{\rm TD}^{\rm LSM}$, 
as $g_Y^{\rm TD}/ g_Y^{\rm SM} =(F_{\rm TD}/F_{\rm TD}^{\rm LSM})^{-1} 
\simeq 0.18$--$0.20$. 
Such a difference may be useful for experimental search for the
techni-dilaton in LHC.

\begin{table}[h]
\begin{center}
\begin{tabular}{|c|c|c|c|} 
\hline 

\hspace{10pt} $N_{\rm TC}$ \hspace{10pt}  
&  \hspace{20pt} $F_{\rm TD}$ [TeV]  \hspace{20pt} 
&  \hspace{20pt} $F_{\rm TD}/F_{\rm TD}^{\rm LSM}$\hspace{20pt} 
&  \hspace{20pt} $g_Y^{\rm TD}/g_{Y}^{\rm SM}$ \hspace{20pt} 
\\
\hline \hline 
2 & 2.42 & 4.91&0.20\\ 
\hline  
 3 & 2.49 &5.07&0.20\\ 
 \hline 
4 & 2.71 &5.52&0.18\\ 
\hline    
\end{tabular} 
\caption{ 
Estimates of the values of $F_{\rm TD}$ 
for each $N_{\rm TC} = 2,3,4$ with $(\Lambda_{\rm ETC}/F_\pi) = 10^4$ and ${\hat S}=0.1$ fixed.  
The values of the ratios $F_{\rm TD}/F_{\rm TD}^{\rm LSM}$ and $g_Y^{\rm TD}/g_{Y}^{\rm SM}$ are also displayed. } 
\label{table:fs} 
\end{center}
\end{table}

\section{Summary and Discussions}

In this paper, 
we have proposed a holographic SWC-TC 
including the bulk flavor/chiral-singlet scalar field $\Phi_X$ 
corresponding to (techni-) gluon condensation,
based on deformation of a hard-wall-type bottom-up holographic QCD
by adjusting the anomalous dimension $\gamma_m$. 
Thanks to the additional explicit bulk scalar field $\Phi_X$, 
we naturally reproduced gluonic $1/Q^4$ terms in the OPE 
of the underlying theory (QCD and SWC-TC) for current correlators, 
in such a way as clearly to distinguish them from the same $1/Q^4$ terms due to 
the chiral condensate in the case of SWC-TC with $\gamma_m \simeq 1$. 
We have analyzed a generic case with $0 \lesssim \gamma_m \lesssim 1$ 
and calculated the masses of the techni-$\rho$ meson ($M_\rho$), 
the techni-$a_1$ meson ($M_{a_1}$), 
and the flavor-singlet scalar meson, techni-dilaton ($M_{\rm TD}$), 
as well as the $S$ parameter.

We have shown that our model with $\gamma_m =0$ and $N_f=3$ 
well reproduces the real-life QCD (See Table~\ref{table:QCD}),
with an indication that the QCD $\rho$ meson mass $M_\rho$ includes 
a (negative) contribution about 10\% from the gluon condensate (See Eq.(\ref{Mrho:G-dep:val})).

For the case of $\gamma_m = 1$, 
we studied the dependence of the $S$ parameter on $(M_\rho/F_\pi)$ 
for several values of the techni-gluon condensation $\Gamma$ (Eq.(\ref{B:def})),    
$\Gamma = 0,\, 5,\, 10$ (Fig.~\ref{S:Bfm:comp}):
$\hat{S}$ decreases monotonically with respect to $(F_\pi/M_\rho)$ to continuously approach zero. 
This implies $(M_\rho/F_\pi)$ necessarily increases when $\hat S$ is required
to be smaller (Eq.(\ref{mf:S})).
It was also shown that, in the region of $\hat S \lesssim 0.1$, 
the techni-gluon condensation reduces the value of $S$ 
maximally about 10\%, 
compared with the previous analysis without techni-gluon contribution~\cite{Haba:2008nz}.

In the generic TC case with $0 \siml \gamma_m \siml 1$, 
we discussed how the meson masses relative to $F_\pi$, 
($M_\rho/F_\pi,\,M_{a_1}/F_\pi,\,M_{\rm TD}/F_\pi$) 
change by varying $\gamma_m$, $\hat{S}$, and 
the techni-gluon condensation $\Gamma$. 
 For fixed value of 
${\hat S}(=0.31)$ (QCD value, Fig.~\ref{mras:B:S3}) 
and $\hat S=0.1$ (minimal requirement for a realistic TC, Fig.~\ref{mras:B}), 
$(M_\rho/F_\pi)$ and $(M_{a_1}/F_\pi)$ are sensitive to neither $\Gamma$ nor 
$\gamma_m$, although the degeneracy between  $(M_\rho/F_\pi)$ and $(M_{a_1}/F_\pi)$ takes place for somewhat larger $\Gamma$.
 In contrast, the techni-dilaton mass has a characteristic feature 
 related to the conformality of SWC-TC: $(M_{\rm TD}/F_\pi)$ substantially 
decreases as $\Gamma$ and/or $\gamma_m$ increases when $\hat{S}$ is fixed.  
Particularly,  $M_{\rm TD}/F_\pi  \to 0$ in the formal limit $\Gamma \to \infty$.

To specify the value of $\Gamma$, 
we considered a couple of typical models of SWC-TC with $\gamma_m \simeq1$ 
and $N_{\rm TC} = 2,3,4$ based on the Caswell-Banks-Zaks infrared fixed point 
in the large $N_f$ QCD. 
Using some specific dynamical features of the conformal anomaly 
indicated by the analysis based on the ladder SD equation,
we found the relation of $\Gamma$ to $(\Lambda_{\rm ETC}/F_\pi)$: 
In the case of $N_{\rm TC}=3$ ($N_f = 4 N_{\rm TC}$) and $S = (N_f/2) \cdot \hat{S} \simeq 0.1$,
we have $\Gamma \simeq 7$ for $(\Lambda_{\rm ETC}/F_\pi) = 10^4$--$10^5$
(required by the FCNC constraint).
Thanks to the large anomalous dimension $\gamma_m$ and 
large techni-gluon condensation $\Gamma$, we had
a relatively light techni-dilaton with mass $M_{\rm TD} \simeq 600 \, \GeV$
compared with $M_\rho \simeq M_{a_1} \simeq 3.8 \, \TeV$ (Table~\ref{table:mass1}).
Note that large values of $M_\rho$ and $M_{a_1}$ are essentially determined 
by the requirement of $S = 0.1$ fairly independently of 
the techni-gluon condensation $\Gamma$, though the degeneracy between them is due to the largeness of $\Gamma$. 
Such large values of $M_\rho$ and $M_{a_1}$ might make the standard signatures through these techni-hadrons  quite invisible at LHC. Note however that the 
largeness of the spectra in our model simply comes from the constraint from the $S$ parameter {\it evaluated by  the TC sector alone}. If we found other effects to reduce the $S$ parameter 
such as the ETC effects implementing the mass of the
SM fermions, we  could pull down the overall scale of the whole techni-hadron spectra, in which case the degenerate techni-$\rho$ and 
techni-$a_1$ as well as the lighter techni-dilaton in our model would have much impact in the LHC phenomenology.

The essential reason for the large $\Gamma$ 
is due to the existence of 
the wide conformal region $F_\pi < \mu < \Lambda_{\rm ETC}$
with $(\Lambda_{\rm ETC}/F_\pi) = 10^4$--$10^5$,
which yields the smallness of the beta function 
through the factor $(\ln{4\Lambda_{\rm ETC}/m})^{-3}$
(see Eq.({\ref{beta})) and hence amplifies 
the techni-gluon condensation in Eq.(\ref{G2:SD})
compared with the ordinary QCD with $\Gamma=1$. 
In the idealized (phenomenologically uninteresting) limit $\Lambda_{\rm ETC}/F_\pi \to \infty$ 
we would have $\Gamma \to \infty$ and hence $M_{\rm TD}/F_\pi  \to 0$.  
The would-be  ``massless'' techni-dilaton is actually decoupled, 
since its decay constant $F_{\rm TD}$ becomes divergent in that limit as is implied in Eq.(4.19).  
There exists no isolated massless techni-dilaton in contrast to 
the chiral symmetry breaking: The scale symmetry is broken both spontaneously and explicitly. 

The predicted mass of holographic techni-dilaton (``conformal Higgs"), 600 GeV,  
lies in the discovery region at LHC. 
The size of Yukawa coupling of the techni-dilaton was also estimated 
through the PCDC relation, which turned out to be somewhat smaller than 
that of the SM Higgs (Table.~\ref{table:fs}). 
More detailed analysis on intrinsic signals at LHC concerning such a 
holographic techni-dilation/conformal Higgs will be explored 
in future publications.
 
\vspace{0.5cm}
Before closing this section,
several comments are in order:
\subsection{Perturbative unitarity-bound and partially EW gauged SWC-TC}

Since the vector meson masses predicted from our analysis 
are of order of a few TeV and hence irrelevant to the unitarity 
(See Table~\ref{table:mass1}), 
it would be the light techni-dilaton that is responsible for   
the perturbative unitarity of $W_LW_L$ scattering. 
The perturbative unitarity-bound on $M_{\rm TD}$ 
can then be estimated through a formula,  
\begin{equation} 
M_{\rm TD} \siml \Lambda_{\rm uni} = \sqrt{8\pi}F_\pi = \sqrt{8 \pi} \cdot \left( \frac{246\, {\rm GeV}}{\sqrt{N_f^{\rm EW}/2}}\right)  
\,, \label{u-formula}
\end{equation}
where $N_f^{\rm EW}$ denotes the number of the techni-fermions charged under the 
electroweak (EW) gauge. 
In the case of the SWC-TC models with $N_f^{\rm EW} =N_f =4N_{\rm TC}$
listed in Table~\ref{table:mass1}, however,
the unitarity bound on the values of $M_{\rm TD}$ may be estimated as 
\begin{equation} 
M_{\rm TD}\Bigg|_{\rm Table~III} 
\lesssim (617, 503 , 436)\, {\rm GeV}\,, 
\qquad {\rm for}\,\, N_{\rm TC}=2,3,4 
\,.
\end{equation} 
Looking at the values of $M_{\rm TD}$ listed in Table~\ref{table:mass1}, 
we see that, for every case of $N_{\rm TC}$, 
some masses of the techni-dilaton are somewhat heavier than that required by    
the unitarity bound above.

The situation with the perturbative unitarity 
would be much improved 
in a class of models (so-called partially gauged model) considered for
other purpose~\cite{Dietrich:2005jn}:
Only a part of techni-fermion flavors
carries the EW charges, while other fermion flavors are EW-singlets
introduced only to achieve the SWC behavior of 
technicolor dynamics. 
Here the idea is to relax our condition 
$N_f^{\rm EW} = N_f = 4 N_{\rm TC}$ such that 
$N_f^{\rm EW} < N_f (= 4N_{\rm TC}) =N_f^{\rm EW} + N_f^{\rm
EW-singlet}$,
which increases $\hat S = S \cdot (2/N_f^{\rm EW})$ 
for the same $S\, (=0.1)$ compared with the analysis 
in Sec.~\ref{ladder:match}.
Actually,
$M_{\rm TD}/\Lambda_{\rm uni} =\frac{1}{\sqrt{8}} (M_{\rm TD}/F_\pi)$
decreases when $\hat S$ increases as shown in Figs.~\ref{mras:ad}, ~\ref{S:Bfm:comp}, 
and Eq.(\ref{mf:S}). 
In Table~\ref{table:mass2} we give two examples of 
the partially gauged models with $N_f = 4 N_{\rm TC}$
(one-doublet models with $N_f^{\rm EW}=2$ and one-family models with
$N_f^{\rm EW}=8$),
where the unitarity condition is fulfilled, $M_{\rm TD}/\Lambda_{\rm uni} \siml 1$.

\begin{table}[h]
\begin{center}
\begin{tabular}{|c|c|c|c|c|} 
\hline 
\multicolumn{5}{|c|}{one-family partially gauged model $N_f^{\rm
 EW} = 8 \leq  4N_{\rm TC}$
$(\Lambda_{\rm uni} = \sqrt{8 \pi}F_\pi \simeq  617\, \GeV)$}\\
\hline
\hspace{10pt} $N_{\rm TC}$ \hspace{10pt}  
& \hspace{10pt} $\log_{10} (\Lambda_{\rm ETC}/F_\pi)$ \hspace{10pt} 
&  \hspace{20pt} $M_{\rm TD}$ [GeV]  \hspace{20pt} 
&  \hspace{20pt} $M_\rho$ [TeV]  \hspace{20pt} 
&  \hspace{20pt} $M_{a_1}$ [TeV]  \hspace{20pt} \\ 
\hline \hline 
2&4 & 777 $\left(^{+106}_{-125}\right)$ &3.75 & 3.82\\ 
\hline 
2&5 &  613 $\left(^{+85}_{-99}\right)$ &3.69  & 3.74\\ 
\hline \hline 
 3&4 &  725 $\left(^{+99}_{-117}\right)$ &3.81 & 3.86\\ 
\hline 
3&5 &  581 $\left(^{+81}_{-94}\right)$ &3.74 & 3.78\\ 
\hline \hline 
4&4 & 686 $\left(^{+94}_{-110}\right)$ &3.84 &3.88\\ 
\hline 
4&5 &  557 $\left(^{+77}_{-90}\right)$ &3.78 & 3.81\\ 
\hline  \hline
\multicolumn{5}{|c|}{one-doublet partially gauged model with $N_f^{\rm
 EW} = 2 < 4N_{\rm TC}$
$(\Lambda_{\rm uni} =\sqrt{8 \pi}F_\pi \simeq  1.23 \, \TeV)$}\\
\hline
  \hspace{10pt} $N_{TC}$ \hspace{10pt}  
& \hspace{10pt} $\log_{10} (\Lambda_{\rm ETC}/F_\pi)$ \hspace{10pt} 
&  \hspace{20pt} $M_{\rm TD}$ [GeV]  \hspace{20pt} 
&  \hspace{20pt} $M_\rho$ [TeV]  \hspace{20pt} 
&  \hspace{20pt} $M_{a_1}$ [TeV]  \hspace{20pt} \\ 
\hline \hline 
2&4 &  863 $\left(^{+119}_{-140}\right)$ &3.57& 3.70\\ 
\hline 
2&5 & 676 $\left(^{+94}_{-110}\right)$ &3.57 & 3.65\\ 
\hline \hline 
 3&4 & 821 $\left(^{+113}_{-133}\right)$ &3.61 & 3.71\\ 
\hline 
3&5 &  645 $\left(^{+90}_{-105}\right)$ &3.59 & 3.71\\ 
\hline \hline 
4&4 & 792 $\left(^{+109}_{-128}\right)$ &3.65 &3.66\\ 
\hline 
4&5 & 623 $\left(^{+87}_{-101}\right)$ &3.62 & 3.68\\ 
\hline    
\end{tabular} 
\caption{ 
Estimates of  the meson masses 
in one-family and one-doublet partially gauged models  
based on the large $N_f$ QCD with $S=0.1$ fixed. 
The values of $M_{\rm TD}$ are estimated 
varying the value of $\kappa$ in the range $0.7 \le \kappa \le 1.3$, 
where the smallest values of $M_{\rm TD}$ correspond to the cases  
with $\kappa=0.7$, while the largest values $\kappa=1.3$. }  
\label{table:mass2} 
\end{center}
\end{table}

\subsection{Possible scaling behaviors of $S$ and 
the conformal phase transition/chiral restoration}

The analyses in this paper were made 
for fixed values of $\hat S$ or $S$ ($\neq 0$)
and $m/\Lambda_{\rm TC} \simeq F_\pi/\Lambda_{\rm TC} = 10^{-4}$--$10^{-5}$. 
The simplest extension of our analysis would imply 
that $S \sim (F_\pi/M_\rho)^2 \to {\rm constant} \neq 0$ as $m/\Lambda_{\rm TC} \to 0$, 
which, as indicated in Table~\ref{table:Mfpilad}, 
seems to be in accord with the straightforward calculation 
based on ladder SD and BS equation~\cite{Harada:2005ru}: 
$\hat S \simeq \,{\rm constant }\,\simeq 0.17$ 
near $\alpha_{*}/\alpha_{\rm cr}=1.13$ 
($F_\pi/\Lambda_{\rm TC} = 10^{-4.6}$) and 
$\hat S \to \,{\rm constant}\, \neq 0$ in the extrapolation toward 
$\alpha_*/\alpha_{\rm cr} \to 1$ ($F_\pi/\Lambda_{\rm TC} \to 0$).

Here we shall mention different possibilities for the scaling of $S$, 
other than ${\hat S} \to $ constant, 
arising as $m/\Lambda_{\rm TC} \to 0$ near the conformal phase transition. 
In the case of conformal phase transition, 
all the dimensionful parameters of order 
${\cal{O}}(m)\, \ll {\cal{O}}(\Lambda_{\rm TC})$ are 
expected to become zero at the critical point, $m/\Lambda_{\rm TC} \to 0$.
Therefore, vector meson mass also goes to zero, $M_\rho/\Lambda_{\rm TC} \to 0$.
Suppose first that $M_\rho/m = {\rm constant}$ in the limit $m/\Lambda_{\rm
TC} \to 0$. Dimensionless parameter $S$ is written in terms of two independent 
dimensionless parameters, say, $\xi$ and $G$, once we fix $\gamma_m$. 
A set of $\xi$ and $G$ is converted to another set of dimensionless 
parameters $M_\rho/m$ and $\Lambda/m$ through Eqs.(\ref{GB:match}) and
(\ref{xi:match}). 
If we fix $M_\rho/m = $ constant, then $S$ is given as a function 
of $\Lambda/m$. 
In Fig.~\ref{S:mr:lamf}, we show the plot of 
$\hat{S}/N_{\rm TC}$ as a function of $\Lambda/m$ for $\gamma_m =1$ and $M_\rho/m =2$. 
Actually, it turns out that $\hat S$ goes to zero with the chiral restoration, 
i.e., $\Lambda/m \to \infty$, whatever the constant value of $M_\rho/m$ is taken to be. 
(If $M_\rho$ were taken to be bigger than $2m$, $\hat S$ would become smaller.) 
This would imply $F_\pi/m \sim F_\pi/M_\rho \to 0$ and hence 
$\condense_m /F_\pi^3 \sim m^3/F_\pi^3 \to \infty$ at the conformal phase transition point. 
This is the behavior somewhat different from that expected from the PS formula~\cite{Haba:2008nz}.

 \begin{figure}
\begin{center}
 \includegraphics[width=9cm]{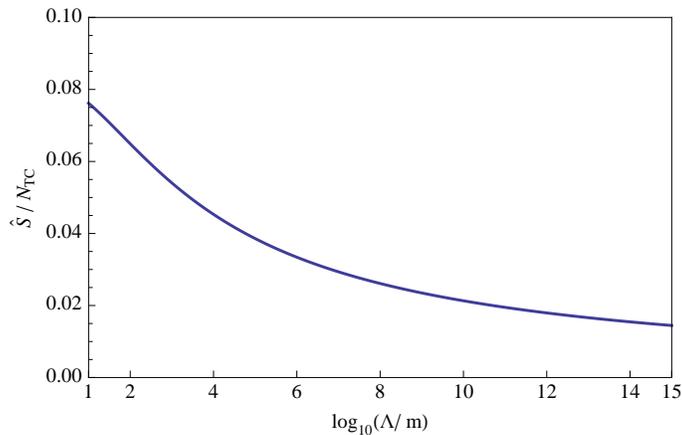}
 \caption{\label{S:mr:lamf}
The plot of $\hat S/N_{\rm TC}$ as a function of $\log_{10}[\Lambda/m]$ 
for $M_\rho/m=2$.} 
\end{center} 
 \end{figure}

}

\section*{Acknowledgments}

We would like to thank M.~Hashimoto and M.~Harada for useful comments and fruitful discussions.  
This work was supported in part by 
the Grant-in-Aid for Nagoya University Global COE Program, ``Quest for Fundamental 
Principles in the Universe: from Particles to the Solar System and the Cosmos'', from the 
Ministry of Education, Culture, Sports, Science and Technology of Japan. 
It was also supported by the JSPS Grant-in-Aid for Scientific Research (S) \#22224003.  
S.M. was supported by the Korean Research Foundation Grant
  funded by the Korean Government
 (KRF-2008-341-C00008).

\appendix 
\renewcommand\theequation{\Alph{section}.\arabic{equation}}

\section{Holographic matching to the operator product expansion of current correlators}  
\label{OPE}

We start with Eq.(\ref{VEOM}), the equation of motion for $V_\mu(q,z)$, 
and rewrite it in terms of $V(q,z)$ defined as $V_\mu(q,z)=\tilde{v}_\mu(q) V(q,z)$ 
as follows: 
\begin{equation} 
  [Q^2 + w^{-1}(z) \partial_z w(z) \partial_z] V(q,z) = 0 
  \,, \label{EOM:V}
\end{equation}
where $Q^2=-q^2$ denotes a Euclidean momentum-squared and 
the induced metric $w(z)$ is given (in the limit $M'=0$) as  
\begin{equation} 
 w(z) = \frac{L}{z} \left(1 + G  \left(\frac{z}{z_m} \right)^4 \right)^2 
\,. \label{omega:def}
\end{equation}
   From Eq.(\ref{GB:sol}) we note that  
\begin{equation} 
  \frac{G}{z_m^4 Q^4} = - \frac{1}{8} \frac{cg_5^4}{L^2} 
\frac{ \langle  \alpha G_{\mu\nu}^2 \rangle}{Q^4} 
\,. 
\end{equation}
 Using this relation, we express the induced metric $w(z)$ as 
\begin{equation} 
  w(z=y/Q) = \frac{QL}{y} \left(  1 - \frac{1}{8} \frac{cg_5^4}{L^2} y^4 
\frac{   \langle  \alpha G_{\mu\nu}^2 \rangle}{Q^4}  \right)^2 
  \,, \label{w:gluon}
\end{equation}  
where the variable $z$ has been replaced with $y=Qz$.

Consider a large Euclidean momentum region $(1/z_m)^2 \ll Q^2 < (1/\ep)^2$. 
 Expanding Eq.(\ref{EOM:V}) in powers of $Q^2$ and using Eq.(\ref{Pi:Vp}), 
we then obtain an asymptotic form of the vector current correlator $\Pi_V(Q^2)$, 
\beqs 
\Pi_{V}(Q^2)\Bigg|_{(1/z_m)^2 \ll Q^2 <(1/\ep)^2} =  Q^2 \left[~\frac{L}{2 g_5^2}\ln{Q^2} 
+ c \frac{2}{3} \frac{g_5^2}{L} \frac{  \langle \alpha G_{\mu\nu}^2 \rangle}{Q^4} 
+ {\cal{O}}(\frac{1}{Q^8})
~\right]
\,. 
\label{PiV:OPE}
\eeqs 
This expression may be compared with the form 
of the operator product expansion (OPE) 
for arbitrary $\gamma_m$, 
\begin{eqnarray}
  \Pi_V \left( Q^2 \right)\Bigg|_{\rm OPE} & = & 
Q^2 \left[\frac{N_{\rm TC}}{24 \pi^2}
\log \left( \frac{Q^2}{\mu^2} \right) 
 - \frac{1}{24 \pi}  \frac{  \langle \alpha G_{\mu\nu}^2 \rangle}{Q^4} 
+ 2 \pi
\frac{ \alpha \condense^2_\mu \mu^{-2\gamma_m}}{Q^{2(3-\gamma_m)}} 
\right] 
\,. 
\label{OPEVA}
\end{eqnarray}
  Then we find the following matching conditions: 
\beqs
\frac{L}{g_5^2} = \frac{N_{\rm TC}}{12 \pi^2}
\,, 
\qquad  
c = -\frac{1}{16 \pi}\frac{L}{g_5^2} = -\frac{N_{\rm TC}}{192 \pi^3} 
\,. 
\eeqs

Similar discussion on the axial-vector current correlator $\Pi_A$ 
provides the same matching conditions for $(L/g_5^2)$ and $c$ 
as it should.

The high-energy expansion form in Eq.(\ref{PiV:OPE}) does not 
include the chiral condensate term behaving as 
$1/Q^{2(3-\gamma_m)}$ as in Eq.(\ref{OPEVA}). 
In order to incorporate this missing term,  
we shall introduce higher dimensional interaction terms constructed from 
the bulk scalar field $\Phi$ and the left- and the right-gauge fields $L_M$ and $R_M$.

We consider the following dimension-six operators 
which are invariant under the five-dimensional $SU(N_f)_L \times SU(N_f)_R$ gauge symmetry: 
\beqs
\Delta {\cal{L}}_5  =\frac{L^2}{ g_5^2}e^{c g_5^2 \Phi_X} \Big(
\frac{C_{LL}}{2}\tr\left[\Phi^\dagger \Phi\right]
\tr\left[{L_{MN}L^{MN}}+{R_{MN}R^{MN}}\right]
+C_{LR} \tr\left[{\Phi^\dagger L_{MN}\Phi R^{MN}}\right]
\Big)
\,, 
\label{S5:int}
\eeqs
where $C_{LL}$ and $C_{LR}$ are the dimensionless coupling constants. 
The Lagrangian $\Delta {\cal L}_5$ gives shifts to the kinetic terms of the 
vector ($V_M\equiv (L_M+R_M)/\sqrt{2}$) and axial-vector ($A_M\equiv(L_M-R_M)/\sqrt{2}$) gauge fields as follows: 
\begin{equation} 
\frac{L^2}{ g_5^2}e^{c g_5^2 \Phi_X} \Big(
\frac{C_{V}}{2}v^2 (z)
\tr\left[V_{MN}V^{MN}\right]
+ \frac{C_{A}}{2}v^2(z)
\tr\left[A_{MN}A^{MN}\right]
\Big)
\,, \label{shift:kin}
\end{equation}
where $C_{V,A} = C_{LL} \pm \frac{1}{2}C_{LR}$.

Let us focus on the vector sector taking $C_A = 0$ for simplicity. 
Then we see that the $C_V$ term in Eq.(\ref{shift:kin}) modifies 
the induced metric $w(z)$ in Eq.(\ref{omega:def}) as 
\begin{eqnarray} 
w(z) \rightarrow 
\tilde{w}(z) 
= 
\frac{L}{z} \left(1+G \left(\frac{z}{z_m}\right)^4 \right)^2 
\left(1 -  C_V L^2 v^2(z) \right)
\,. \label{w:tilde}
\end{eqnarray} 
  It should be noted from Eqs.(\ref{condense:sol}) and 
(\ref{v:sol})-(\ref{c2:v}) 
that $v^2(z)$ in Eq.(\ref{w:tilde}) is expressed 
(in the chiral limit $M=0$) as
\begin{equation} 
 v^2(z=y/Q)  \simeq \frac{y^{2(3-\gamma_m)}}{3(3-\gamma_m)^2} 
\left( \frac{g_5^2}{L}  \right)^2 \frac{\condense_{1/L}^2 L^{2\gamma_m-2}}{Q^{2(3-\gamma_m)}}
\,, 
\end{equation}  
  where we have neglected higher order terms in $1/Q^2$ expansion. 
Thus we see that the modified-induced metric $\tilde{w}(z)$ now 
includes the desired $\bar TT$ term:  
\beqs
\tilde{w}(z=y/Q) 
\ni 
\frac{QL}{y} 
\left[ - C_V \left(\frac{g_5^2}{L}\right)^2\frac{y^{2(3-\gamma_m)} }{3(3-\gamma_m)^2} 
\frac{L^{2\gamma_m}\condense^2_{1/L}}{Q^{2(3-\gamma_m)}} 
\right] 
\,. 
\eeqs 
  Combining this with Eq.(\ref{w:gluon}), 
we obtain the total expression of the high-energy expansion for 
$\Pi_V$: 
\beqs
\Pi_{V}(Q^2) \Bigg|_{(1/z_m)^2 \ll Q^2 < (1/\ep)^2} 
= Q^2 \left[~\frac{L}{2 g_5^2}\ln{Q^2} 
+ c \frac{2}{3} \frac{g_5^2}{L} \frac{\langle \alpha G_{\mu\nu}^2 \rangle}{Q^4} 
+ C_6^{V} L^{2 \gamma_m} \frac{\condense^2_{1/L}}{Q^{2(3-\gamma_m)}}
+{\cal{O}}(\frac{1}{Q^8})
~\right]
\,,
\label{PiV:complete3}
\eeqs 
where 
\beqs
C_6^V = C_V 
\frac{\sqrt{\pi}}{6}\frac{(\Gamma(2-\gamma_m))^3}
{\Gamma\left(\frac{7}{2}-\gamma_m \right)}
\frac{g_5^2}{L}
\,. 
\label{C6}
\eeqs 
Comparing Eq.(\ref{PiV:complete3}) with the $\langle \bar{T} T \rangle^2$ term 
in Eq.(\ref{OPEVA}), we determine the coefficient $C_V$ as 
\beqs
C_V 
= 12 \sqrt{\pi} \alpha 
\frac{\Gamma\left(\frac{7}{2} - \gamma_m \right)}{(\Gamma(3-\gamma_m))^3}
\,. 
\frac{L}{g_5^2}
\,.
\eeqs

For $C_A \neq 0$, one can similarly perform the high-energy expansion of $\Pi_A$.  
As a consequence of matching with the OPE for $\Pi_A$, 
we get 
\begin{equation} 
 C_A = - 12 \sqrt{\pi} \alpha \frac{\Gamma(\frac{7}{2} - \gamma_m)}{(\Gamma(3-\gamma_m))^3} 
 \frac{L}{g_5^2} + \frac{1}{(3-\gamma_m)(2-\gamma_m)}
\,. 
\end{equation}

Thus, it is shown that the present model with the higher dimensional 
terms in Eq.(\ref{S5:int}) added completely reproduces the OPE for the 
current correlators up to terms suppressed by $1/Q^8$.

\section{The Pagels-Stokar formula and chiral condensate} 
\label{PS:cond}

\subsection{Relationship between decay constant and dynamical fermion mass}

The Pagels-Stokar formula relates 
the mass function $\Sigma(Q^2)$ with $F_\pi$ as 
\beqs
\label{PS:int}
F_\pi^2 = \frac{N_{\rm TC}}{4\pi^2} m^2_{\rm PS} 
\int_{0}^{(\Lambda^2/m_{\rm PS}^2 \to \infty)} d x x \frac{\Sigma^2(x)-\frac{x}{4}\frac{d}{d
x}\Sigma^2(x)}
{(x+\Sigma^2(x))^2},
\eeqs
where $\Sigma(x) = \Sigma(Q^2)/m_{\rm PS}$ which may be parametrized as 
\begin{eqnarray}
\label{sigma:para}
\Sigma(x) =\Bigg\{
\begin{array}{ll}
x^{\frac{\gamma_m}{2}-1} & \hspace{10pt} {\rm for } \,\,\, x >1 \\
1 & \hspace{10pt} {\rm for }\,\,\,x <1 
\end{array}
\,. 
\end{eqnarray}
By using Eq.(\ref{sigma:para}), Eq.(\ref{PS:int}) is calculated as~\footnote{
One may neglect $\Sigma^2(x) = x^{\gamma_m-2}$ in the denominator of 
the integrand of Eq.(\ref{PS:int}) since in the integration 
dominant contributions come from the UV region where 
$\Sigma^2(x) = x^{\gamma_m-2} \ll x$ for $0 \lesssim \gamma_m \lesssim 1$. 
Then the form of Eq.(\ref{PS:coe}) would be changed to  
$
\frac{4\pi^2}{N_{\rm TC} } \frac{F_\pi^2}{m^2_{\rm PS}} 
\simeq  
-\frac{1}{2}+\ln{2} 
+ \frac{1}{2}\frac{\left(3-\frac{\gamma_m}{2}\right)}{(2-\gamma_m)}$. 
} 
\beqs
\label{PS:coe}
\frac{4\pi^2}{N_{\rm TC} } \frac{F_\pi^2}{m^2_{\rm PS}} 
&\simeq &
\left(\int_{0}^{1} d x  \frac{x}{(x+1)^2}
+\frac{\left(3-\frac{\gamma_m}{2}\right)}{2}
\int_{1}^{\infty} d x  \frac{x^{\gamma_m-1}}
{(x+x^{\gamma_m-2})^2}\right) \nonumber \\
&=&  -\frac{1}{2}+\ln{2} + \frac{1}{2(3-\gamma_m)^2}
\left((3-\gamma_m) + H_{\frac{-1}{2(3-\gamma)}}-
H_{\frac{-(4-\gamma)}{2(3-\gamma)}}\right),
\eeqs
where $H_n$ is harmonic number with order of $n$ defined as 
$H_{n}=\sum_{k=1}^{n}1/k^n$. 
   For $\gamma_m = 0,1$, 
Eq.(\ref{PS:coe}) is evaluated as 
\beqs
\label{PS:coe2}
\frac{4\pi^2}{N_{\rm TC} } \frac{F_\pi^2}{m^2_{\rm PS}} 
&\simeq& \Bigg\{
\begin{array}{ll}
0.63  & \hspace{10pt}{\rm for } \,\,\, \gamma_m =0 \\
1.00  & \hspace{10pt}{\rm for } \,\,\, \gamma_m = 1 \\ 
\end{array}  
\,, 
\eeqs
and for $\gamma_m=2$ we have 
$\frac{4\pi^2}{N_{\rm TC} } \frac{F_\pi^2}{m^2_{\rm PS}} \simeq \ln{(\Lambda/m)^2}$.

\subsection{Relationship between chiral condensate and dynamical fermion mass}

The chiral condensate $\condense$ evaluated at a UV scale $\Lambda$ 
is given as 
\beq
\label{cond:int}
\condense_{\Lambda} = \frac{N_{\rm TC}}{4\pi^2} m^3 
\int_{0}^{(\Lambda^2/m^2 \to \infty)} d x \frac{x \Sigma(x)}
{x+\Sigma^2(x)}.
\eeq
One can calculate Eq.(\ref{cond:int}) using 
the following relation derived based on analysis of 
the ladder SD equation~\cite{Bando:1987we}~\footnote{Neglecting 
the $\Sigma(x)$ in the denominator, which is not dominant
in the integral, and substituting Eq.(\ref{sigma:para}) into Eq.(\ref{cond:int}), 
we may evaluate Eq.(\ref{cond:int}) to reach a form, 
\begin{equation}
\label{cond:loop}
-\frac{4\pi^2 }{N_{\rm TC}} \frac{\condense_\Lambda}{m^3} 
\simeq \left[1-\ln{2} + \frac{2}{\gamma_m} \left( 
\left(\frac{\Lambda}{m}\right)^{\gamma_m}-1 \right) \right]
\,, 
\end{equation} 
which turns out to be good approximation and gives the same numbers 
as those in Eqs.(\ref{cond:WTC}) and (\ref{cond:NJL}), 
except for the QCD case with $\gamma_m=0$. }  
\beq
\label{cond:ladder}
\int_{0}^{(\Lambda^2/m^2 \to \infty)} d x \frac{x \Sigma(x)}
{x+\Sigma^2(x)} 
=
\frac{\frac{d}{d x}\Sigma(x)}{\frac{d}{d x}(\lambda(x)/x)} \Bigg|_{x=(\Lambda/m)^2}
\,, 
\eeq 
where $\lambda(x) = \frac{1}{4}\frac{\alpha}{\alpha_{\rm cr}}$ with
$\alpha_{\rm cr} = \pi/(3 C_2 (F))$.

In the case of SWC-TC with $\gamma_m \simeq 1$ 
where $\alpha (\simeq \alpha_*) \simeq \alpha_{\rm cr}$, 
we have $\lambda = 1/4$. 
Using Eqs.(\ref{sigma:para}) and (\ref{cond:ladder})
we then evaluate Eq.(\ref{cond:int}) as   
\beq
-\frac{4\pi^2 }{N_{\rm TC}} \frac{\condense_{\Lambda}}{m^3}\Bigg|_{\rm SWC-TC}  
= 2 \cdot Z_m^{-1} 
\,, 
\label{cond:WTC}
\eeq
where $Z_m = (\Lambda/m)^{-1}$ is the mass renormalization constant.

In the case of QCD with $\gamma_m \simeq 0$, 
$\Sigma(x)$ and $\lambda(x)$ are expressed as  
\beqs
\label{sigma:QCD}
\Sigma(x)&=& \frac{1}{x}\left[1+ \frac{1}{2A} \ln x\right]^{A/2-1} 
\,, \\
\lambda(x)& =& \frac{A/2}{\ln{x} + 2A}
\,, \label{lambda:QCD}
\eeqs
where $A = 1/(b \alpha_{\rm cr})$ 
with $b$ being the coefficient of QCD beta function at one-loop order,  
$b = \frac{1}{6\pi}\left(11 N_c -2N_f\right)$. 
Substituting Eqs.(\ref{sigma:QCD}) and (\ref{lambda:QCD}) 
into Eq.(\ref{cond:ladder}), we have
\begin{eqnarray} 
 \label{cond:QCD}
-\frac{4\pi^2 }{N_c} \frac{\condense_{\Lambda}}{m^3} \Bigg|_{\rm QCD}
&\simeq& \frac{4}{(2 A)^{A/2}} \cdot Z_m^{-1} 
\nonumber \\ 
&\simeq &
 Z_m^{-1}  \cdot 
\Bigg\{  
\begin{array}{ll} 
3.1 & \hspace{10pt} \textrm{for real life QCD with } N_c=N_f=3 \, (A=8/9)\\ 
3.3 & \hspace{10pt} \textrm{for large } N_c \textrm{ QCD} \, (A=9/11)
\end{array}
\,, 
\end{eqnarray}
where $Z_m = (\ln (\Lambda^2/m^2))^{-A/2}$. 

   For the case of constant mass $\Sigma(x)=1$ which corresponds to the case with 
$\gamma_m=2$ as in the Nambu-Jona-Lasinio (NJL) model, 
we can straightforwardly calculate Eq.(\ref{cond:int}) to get 
\begin{equation} 
-\frac{4\pi^2 }{N_{\rm TC}} \frac{\condense_{\Lambda}}{m^3} \Bigg|_{\rm NJL} 
= 1 \cdot Z_m^{-1} 
\,, \label{cond:NJL}
\end{equation} 
where $Z_m=(\Lambda^2/m^2)^{-1}$

Finally using $\condense_m = Z_m \condense_{\Lambda}$, 
we thus reach a result 
\beqs
-\frac{4\pi^2}{N_{\rm TC}} \frac{\condense_m}{m^3} 
&\simeq& 
\Bigg\{
\begin{array}{lll}
3 & \hspace{10pt} {\rm for } & \gamma_m =0 \\
2 & \hspace{10pt} {\rm for } & \gamma_m = 1 \\
1 & \hspace{10pt} {\rm for } & \gamma_m = 2 
\end{array}
\,. 
\end{eqnarray}


\end{document}